\documentclass[aps,prd,superscriptaddress,eqsecnum,showpacs,floatfix]{revtex4}
\pdfoutput=1
\usepackage{amssymb,amsmath,comment,color,slashed,footnote,feynmf,mathrsfs}
\usepackage{graphicx}
\usepackage{color}
\usepackage{amssymb}
\usepackage[colorlinks,urlcolor=blue,citecolor=blue]{hyperref}
\def\hhref#1{\href{http://arxiv.org/abs/#1}{arXiv:#1}} 

\begin{document}

\title{Fractionalized Non-Self-Dual Solutions in the ${\mathbb C}{\mathbb P}^{N-1}$ Model}
\author{Robert Dabrowski and Gerald V. Dunne}
\affiliation{Physics Department, University of Connecticut, Storrs CT 06269, USA}

\begin{abstract}
We study non-self-dual classical solutions in the ${\mathbb C}{\mathbb P}^{N-1}$ model with ${\mathbb Z}_N$ twisted boundary conditions on the spatially compactified cylinder. These solutions have finite, and fractional, classical action and topological charge, and are `unstable' in the sense that the corresponding fluctuation operator has negative modes. We propose a physical interpretation of these solutions as saddle point configurations whose contributions to a resurgent semi-classical analysis of the quantum path integral are imaginary non-perturbative terms which must be cancelled by infrared renormalon terms generated in the perturbative sector.

\end{abstract}
                                          
\maketitle

\section{Introduction}

Recent work has emphasized the physical significance of  ``bions'', topologically trivial vacuum configurations that are locally molecules of instantons and anti-instantons,  for the  study of confinement and chiral symmetry breaking in QCD and supersymmetric gauge theory \cite{Shifman:2008ja,Poppitz:2009uq,Poppitz:2012sw,Shuryak:2012aa,Anber:2011de,Unsal:2012zj}. These are extensions of important related early work by Yung \cite{Yung:1987zp}, and Rubakov and Shevkov \cite{Rubakov:1994qn}.  Using spatial compactification and the principle of continuity,  in gauge theories and ${\mathbb C}{\mathbb P}^{N-1}$ models a correspondence has been demonstrated between infrared renormalons  and certain fractionalized non-perturbative bion (and bion-molecule) objects \cite{Argyres:2012vv,Argyres:2012ka,Dunne:2012ae,Dunne:2012zk}. Motivated by these results, in this paper we study non-self-dual classical solutions of the ${\mathbb C}{\mathbb P}^{N-1}$ model with twisted boundary conditions on the spatially compactified cylinder. These non-self-dual solutions are solutions to the second-order classical equations of motion, but are not solutions to the first-order instanton equations. They have finite action, but are `unstable' in the sense that the fluctuation operator around these classical solutions has negative modes, and so these solutions are saddle-points of the action rather than minima. They were found and classified by Din and Zakrzewski \cite{Din:1980jg,zbook} for ${\mathbb C}{\mathbb P}^{N-1}$ on ${\mathbb R}^2$ and ${\mathbb S}^2$. Here we investigate these solutions on the spatially compactified cylinder, ${\mathbb S}^1_L \times{\mathbb R}^1$, with ${\mathbb Z}_N$ twisted boundary conditions, and show that the non-self-dual solutions fractionalize with a rich pattern of actions and charges, that can be identified locally with fractionalized instantons that occur in twisted ${\mathbb C}{\mathbb P}^{N-1}$ models \cite{Bruckmann:2007zh,Brendel:2009mp,Shifman:2008kj}.

Our motivation is to propose a new physical interpretation of these `unstable' finite action classical solutions, in light of recent work on the ${\mathbb C}{\mathbb P}^{N-1}$ model using resurgent asymptotic analysis  \cite{Dunne:2012ae,Dunne:2012zk}, in which the perturbative infrared renormalons of ${\mathbb C}{\mathbb P}^{N-1}$ were identified with fractionalized multi-instanton configurations [instanton--anti-instanton bions and bion-molecules] in the non-perturbative sector. This identification relies crucially on the spatial compactification, which regularizes the otherwise-ill-defined (due to the instanton scale modulus problem) non-perturbative instanton gas description, and generates ${\mathbb Z}_N$ twisted boundary conditions, which in turn lead to the appearance of fractionalized instanton configurations. Certain multi-instanton amplitudes produce imaginary non-perturbative contributions which were shown to  cancel against terms produced by the analysis of the non-Borel-summable (due to infrared renormalons) perturbative sector. Taken together, as a resurgent semi-classical expansion, the imaginary ambiguities in the perturbative and non-perturbative sectors cancel, rendering the theory fully self-consistent. This is a concrete field theoretic realization of the Bogomolny-Zinn-Justin (BZJ) cancellation mechanism of quantum mechanics \cite{Bogomolny:1980ur,ZinnJustin:1981dx,Balitsky:1985in}.

The analysis of ${\mathbb C}{\mathbb P}^{N-1}$ bion amplitudes in \cite{Dunne:2012ae,Dunne:2012zk}, and in the related Yang-Mills studies in \cite{Argyres:2012vv,Argyres:2012ka}, was based on the standard instanton calculus approach that considers the interactions amongst the constituents of  classical configurations consisting of far-separated instantons and anti-instantons \cite{Vainshtein:1981wh,Yung:1987zp,Schafer:1996wv,Dorey:2002ik,Vandoren:2008xg}. These bions and bion-molecules are {\it approximate} classical solutions, and for certain alignments and fermion content, the bions or bion-molecules have unstable negative modes leading to imaginary non-perturbative contributions  \cite{Argyres:2012vv,Argyres:2012ka,Dunne:2012ae,Dunne:2012zk}. However, we point out here that in precisely these two asymptotically free quantum field theories, 4d Yang-Mills theory and 2d ${\mathbb C}{\mathbb P}^{N-1}$,  there exist {\it exact} non-self-dual solutions, consisting locally of combinations of instantons and anti-instantons. These classical solutions have finite action, but have negative fluctuation modes. For 4d Yang-Mills theory, there is a mathematical existence proof for these non-self-dual solutions in $su(2)$ \cite{sibner}, explicit ansatz forms \cite{Sadun:1992vj,Burzlaff:1980hr,Schiff:1991bu}, and simple embedding constructions for $su(N)$ with $N\geq 4$ \cite{Vandoren:2008xg}, but these Yang-Mills solutions are somewhat unwieldy. On the other hand, for ${\mathbb C}{\mathbb P}^{N-1}$ there is a simple construction for generating these solutions on ${\mathbb R}^2$ and ${\mathbb S}^2$ \cite{Din:1980jg,zbook}, which makes them easy to analyze. While a number of mathematical properties of these non-self-dual solutions have been studied \cite{Din:1980jg,zbook,Jack:1981rq}, no concrete physical interpretation has been proposed. Motivated by the above discussion of resurgent analysis of 4d Yang-Mills theory and 2d ${\mathbb C}{\mathbb P}^{N-1}$  \cite{Argyres:2012vv,Argyres:2012ka,Dunne:2012ae,Dunne:2012zk}, where spatial compactification and ${\mathbb Z}_N$ twisted boundary conditions play  key roles, in this paper we study the unstable non-self-dual classical solutions in ${\mathbb C}{\mathbb P}^{N-1}$ with twisted boundary conditions. The effect of twisted boundary conditions on self-dual instanton solutions has been studied in detail previously, for ${\mathbb C}{\mathbb P}^{N-1}$ \cite{Bruckmann:2007zh,Brendel:2009mp} and Yang-Mills \cite{Lee:1997vp,Kraan:1998pm}. While the physical interpretation of these caloron solutions  is quite different  \cite{Dunne:2012ae,Dunne:2012zk}, many technical details are similar. 

In this paper we generalize the work of Din and Zakrzewski on non-self-dual solutions to incorporate twisted boundary conditions, and show that the solutions persist, and lead to a rich structure of fractionalized topological charges. Our ultimate motivation is to identify these exact saddle-point solutions with a resurgent trans-series  expansion of the field theoretic path integral at weak coupling:
\begin{eqnarray}
\int {\mathcal D}n\, e^{-\frac{1}{g^2} S[n]} = \sum_k \sum_l \sum_p c_{k,l,p} \,  e^{-k/g^2}\, g^{2l}\, \left(\ln \left(-\frac{1}{g^2}\right)\right)^p
\label{trans}
\end{eqnarray}
Here the sum over $k$ covers all multi-instanton sectors, the sum over $l$ covers perturbation theory and all perturbative fluctuations about each multi-instanton sector, and the log sum encapsulates quasi-zero mode contributions. This trans-series structure arises generically from a full semi-classical expansion around all critical points, both minima (instantons) and saddle points (non-self-dual classical solutions).
While the dominant non-perturbative contributions for a given topological charge come from instantons, the non-self-dual classical solutions are saddle points, so they produce higher-order contributions. Nevertheless, the results of \cite{Dunne:2012ae,Dunne:2012zk} show that these saddle point contributions should be included for the semi-classical trans-series expansion (\ref{trans}) to be fully self-consistent.
This is because, due to the appearance of negative fluctuation modes, these contributions will generically be complex, and for consistency of the theory they must be canceled by imaginary non-perturbative contributions arising from the non-Borel-summable nature of the perturbative expansions about the vacuum and each instanton sector. ``Resurgence'' is the statement that these cancellations occur to all orders in the expansion (\ref{trans}), and this has been demonstrated explicitly for low orders in ${\mathbb C}{\mathbb P}^{N-1}$ models  \cite{Dunne:2012ae,Dunne:2012zk}.

\section{Classical solutions of $\mathbb{CP}^{N-1}$}
\label{sec:classical}
We begin with a brief review of notation and previous results \cite{Din:1980jg,zbook}.

\subsection{Action and Topological Charge}

The $\mathbb{CP}^{N-1}$ model has classical action
\begin{eqnarray}
S[n] = \int d^2x \left( D_\mu n \right)^\dag \! \left( D_\mu n \right)
\label{action}
\end{eqnarray}
where $n$ is a complex $N$-component vector satisfying $n^\dagger n=1$. The $\mathbb{CP}^{N-1}$ model has a global $U(N)$ symmetry and a local $U(1)$ gauge symmetry, for which the covariant derivative is $D_\mu = \partial_\mu - i A_\mu$, with $A_\mu = -i\, n^{\dag} \partial_\mu n$. 
The 2d manifold over which the integral in (\ref{action}) is taken, and associated boundary conditions, will be specified below. The cases of interest here are ${\mathbb R}^2$, ${\mathbb S}^2$ and ${\mathbb S}^1_L \times{\mathbb R}^1$. With a Bogomolny factorization, the action can be re-written
\begin{align}
S = \int d^2 x \left[ \frac{1}{2} \Big| D_\mu n \pm i \epsilon_{\mu \nu} D_\nu n \, \Big|^2 \mp i \epsilon_{\mu \nu} \left(D_\nu n \right)^\dag \! D_\mu n \right]
\label{bog}
\end{align}
from which we identify the topological charge
\begin{eqnarray}
Q= \int d^2x \, i \epsilon_{\mu \nu} \left(D_\nu n \right)^\dag \! D_\mu n = \int d^2x \, \epsilon_{\mu \nu} \partial_\mu A_{\nu} 
\label{charge}
\end{eqnarray}
Thus, $S\geq |Q|$, and we note that for finite action solutions on ${\mathbb R}^2$ and ${\mathbb S}^2$, $Q$ is an integer multiple of $2\pi$.

Another useful representation of the $\mathbb{CP}^{N-1}$ model is in terms of the $N\times N$ holomorphic projector field, $
\mathbb{P} \equiv  n\, n^{ \, \dag}
$, which satisfies $\mathbb{P}^2 = \mathbb{P} = \mathbb{P}^\dag $, and $\text{Tr}\, \mathbb{P}= 1$. The action (\ref{action}) and topological charge (\ref{charge}) take the simple form
\begin{eqnarray}
S &=& 2 \int d^2 x \, \text{Tr}\left[ \partial_{z} \mathbb{P} \, \partial_{\bar{z}} \mathbb{P} \right]\label{sqp1}\\
Q &=& 2\int d^2x \, \text{Tr}\Big[ \mathbb{P} \, \partial_{\bar{z}} \mathbb{P} \, \partial_{z} \mathbb{P} - \mathbb{P} \, \partial_{z} \mathbb{P} \,\partial_{\bar{z}} \mathbb{P} \Big]
\label{sqp2}
\end{eqnarray}
where $z=x_1+ix_2$.
This projector representation is particularly convenient for analyzing non-self-dual solutions.

\subsection{Self-dual (instanton) solutions}

From the Bogomolny factorization (\ref{bog}), we deduce the  first-order instanton (self-duality) equations:
\begin{align}
D_\mu n = {} \pm i \epsilon_{\mu \nu} D_\nu n
\label{first}
\end{align}
Explicit instanton solutions are simple to construct using the homogeneous field $\omega$, where $n\equiv \omega/|\omega |$, in terms of which the first-order instanton equations reduce to the Cauchy-Riemann equations, so that instantons correspond to holomorphic vectors, $\omega=\omega(z)$, and anti-instantons correspond to anti-holomorphic vectors, $\omega=\omega(\bar z)$. In the projector representation, the instanton equations are:
\begin{eqnarray}
\partial_{\bar{z}} \mathbb{P} \, \mathbb{P} = 0 \quad ({\rm instanton})\quad, \quad
\partial_{z} \mathbb{P} \, \mathbb{P} = 0 \quad (\text{anti-instanton})
\label{pinst}
\end{eqnarray}
The instanton equations are solved by the $N\times N$ holomorphic projectors, ${\mathbb P}=\frac{\omega\, \omega^\dagger}{\omega^\dagger \omega}$, with $\omega=\omega(z)$.

\subsection{Non-self-dual solutions}

The critical points of the action (\ref{action}) are solutions to 
the full (second-order) classical equations of motion:
\begin{eqnarray}
D_\mu D_\mu n - (n^\dag \! \cdot \! D_\mu D_\mu n) \, n = 0 \qquad {\rm or}\qquad \left[ \partial_z \partial_{\bar z} \mathbb{P} \,,\, \mathbb{P} \right] = 0
\label{second}
\end{eqnarray}
Note that solutions to the instanton equations (\ref{first}) or (\ref{pinst}) are automatically solutions to (\ref{second}), but not vice versa. 

Explicit non-self-dual solutions can be generated from an initial self-dual (instanton) solution by the following procedure of projection operations \cite{Din:1980jg,zbook}.
We define the projection operator $Z_+$ acting on a classical solution $\omega(z, \bar z)$ as:
\begin{align}
Z_+:\omega\to Z_+ \omega \equiv \partial_z \, \omega - \frac{\left( \omega^\dag \, \partial_z \, \omega \right)}{\omega^\dagger \omega}\,\omega \qquad, \qquad Z_+:n\to Z_+ n\equiv \frac{Z_+\omega}{|Z_+ \omega|}
\label{zplus}
\end{align}
It is straightforward to verify using elementary identities that if $\omega$ is a classical solution, then $Z_+\omega$ is also a classical solution \cite{Din:1980jg,zbook}. We can therefore generate a  tower of classical solutions by starting with an initial instanton configuration, $\omega = \omega_{(0)}(z)$, and repeatedly acting with $Z_+$:
\begin{eqnarray}
\omega_{(k)}(z, \bar z)\equiv Z_+^k \omega_{(0)}(z)
\label{tower}
\end{eqnarray}
Notice that the projection operation (\ref{zplus}) introduces dependence on $\bar z$, due to the adjoint operation, so the projected solutions are no longer instantons. Nevertheless, they satisfy the second-order classical equations of motion. Moreover, the tower of projection operations eventually truncates, after at most $(N-1)$ steps in $\mathbb{CP}^{N-1}$, because eventually the classical solution becomes an anti-instanton, for which $Z_+\omega(\bar z)=0$. (Indeed, we could have begun with an anti-instanton and projected up the ladder in the other direction; this is equivalent.) Din and Zakrzewski proved that on ${\mathbb R}^2$ and ${\mathbb S}^2$, this  repeated projection operation (\ref{tower}) produces {\it all}  finite action non-self-dual classical solutions \cite{Din:1980jg,zbook}:
\begin{align}
\omega_{(0)} \;
\xrightarrow[]{\;\; Z_+ \;\;} \; 
\omega_{(1)} \;
\xrightarrow[]{\;\; Z_+ \;\;} \;\
\cdots \;
\xrightarrow[]{\;\; Z_+ \;\;} \;
\omega_{(k)} \;
\xrightarrow[]{\;\; Z_+ \;\;} \;
\cdots \;
\xrightarrow[]{\;\; Z_+ \;\;} \;
\omega_{(N-1)} \;
\xrightarrow[]{\;\; Z_+ \;\;}
0
\label{ZakMappingInstantonTrain}
\end{align}
In the tower (\ref{ZakMappingInstantonTrain}), the initial solution $\omega_{(0)}$ is an instanton, while the final solution $\omega_{(N-1)}$ is an anti-instanton. Note in particular that for ${\mathbb C}{\mathbb P}^{1}$ we do not generate any non-self-dual solutions, as the initial instanton maps directly to an anti-instanton. Thus, we need to consider at least the $N=3$ case: ${\mathbb C}{\mathbb P}^{2}$. Explicit examples are presented below.

\subsection{Action and Topological Charge  of Non-Self-Dual Classical Solutions}

The projector representation is particularly convenient for describing the action and topological charge of the non-self-dual solutions. The solution $\omega_{(k)}$ has action $S_{(k)}$ and topological charge $Q_{(k)}$ given by expressions (\ref{sqp1}, \ref{sqp2}) evaluated on the 
projector 
\begin{eqnarray}
{\mathbb P}_{(k)}\equiv \frac{\omega_{(k)}\, \omega^\dagger_{(k)}}{\omega^\dagger_{(k)} \omega_{(k)}}
\label{pk}
\end{eqnarray}
Using basic algebraic identities and the result \cite{zbook} that for all $k$:
\begin{eqnarray}
{\mathbb P}_{(k)}\bar\partial {\mathbb P}_{(k)}=\sum_{j=0}^{k} \bar\partial {\mathbb P}_{(j)}
\label{id1}
\end{eqnarray}
one can show that:
\begin{eqnarray}
S_{(k)}=Q_{(k)}+2\sum_{j=0}^{k-1} Q_{(j)}
\label{sumid}
\end{eqnarray}
Some useful related identities are listed in the Appendix.
Since the final solution is an anti-instanton, $S_{(N-1)}=-Q_{(N-1)}$, and therefore we see that $S_{(N-1)}=\sum_{j=0}^{N-2} Q_{(j)}$. For example, for ${\mathbb C}{\mathbb P}^{2}$ ($N=3$) we have
\begin{eqnarray}
S_{(0)}= Q_{(0)}\qquad, \qquad 
S_{(1)}=2Q_{(0)}+Q_{(1)}\qquad, \qquad 
S_{(2)}= Q_{(0)}+Q_{(1)}
\label{n3}
\end{eqnarray}
In particular, if the intermediate non-self-dual solution has $Q_{(1)}=0$, then $S_{(1)}=2Q_{(0)}$, and $Q_{(2)}=-Q_{(0)}$. For ${\mathbb C}{\mathbb P}^{N-1}$ on ${\mathbb R}^2$ and ${\mathbb S}^2$, all $S_{(k)}$  and $Q_{(k)}$ are integer multiples of $2\pi$. We show below that with twisted boundary conditions on the spatially compactified cylinder, ${\mathbb S}^1_L \times{\mathbb R}^1$, there is a much richer set of actions and charges.

\subsection{Fluctuation Modes}

\begin{figure}[ht!]
\centering
\includegraphics[width=9cm]{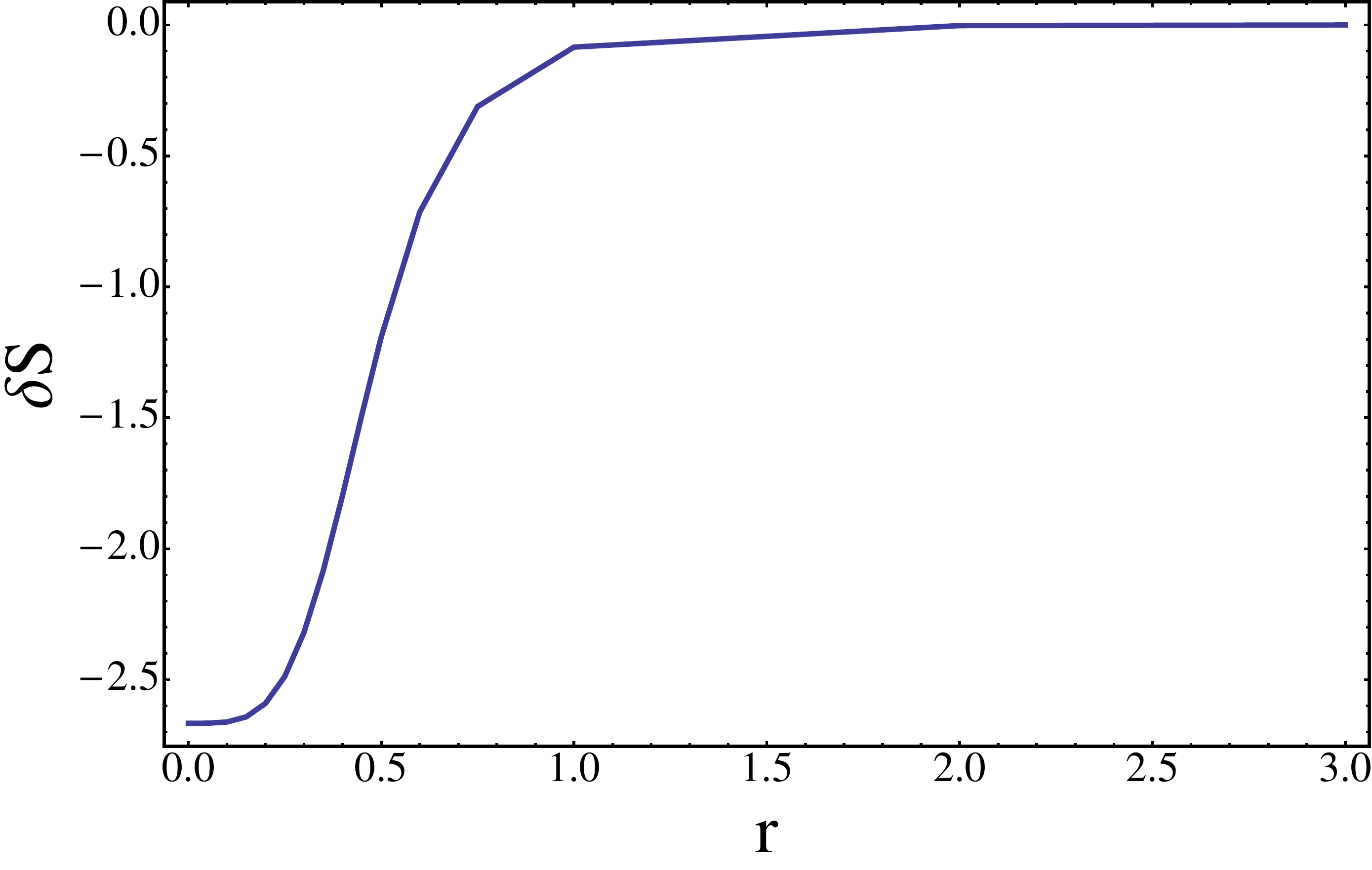}
\caption{The change (\ref{unstable}) in the action under the fluctuation (\ref{change}), for the $Q=0$ non-self-dual configuration plotted below in the second row of Figure \ref{fig:CP2_R2}. The horizontal axis denotes the (symmetric) distance of each object from the center. Notice that at large separation this fluctuation is a zero mode, while at finite separation it becomes a negative mode.}
\label{fig:potential}
\end{figure}

\begin{figure}[htb]
\begin{tabular}{cc}
\includegraphics[width=8cm]{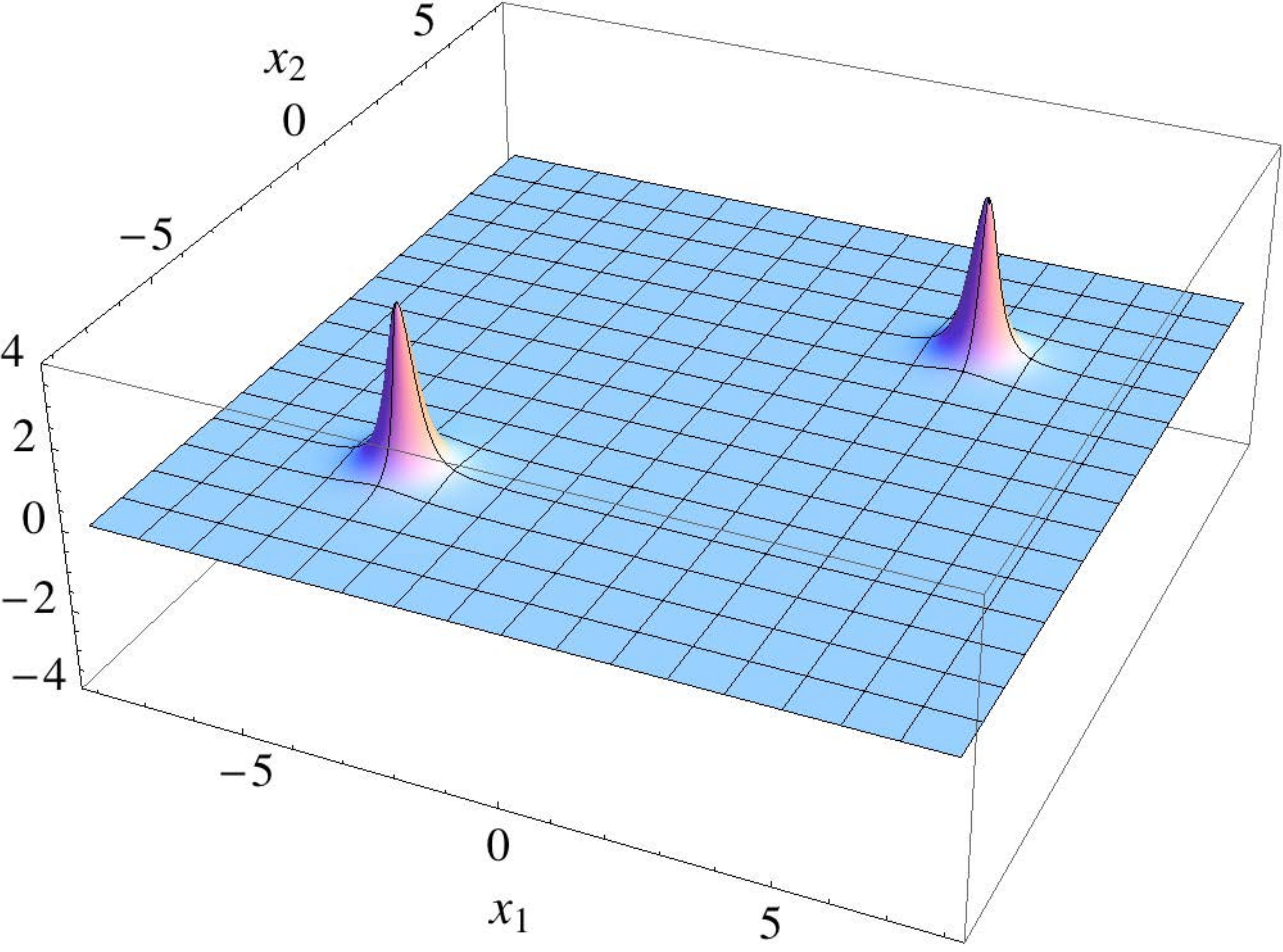} 
 & 
 \includegraphics[width=8cm]{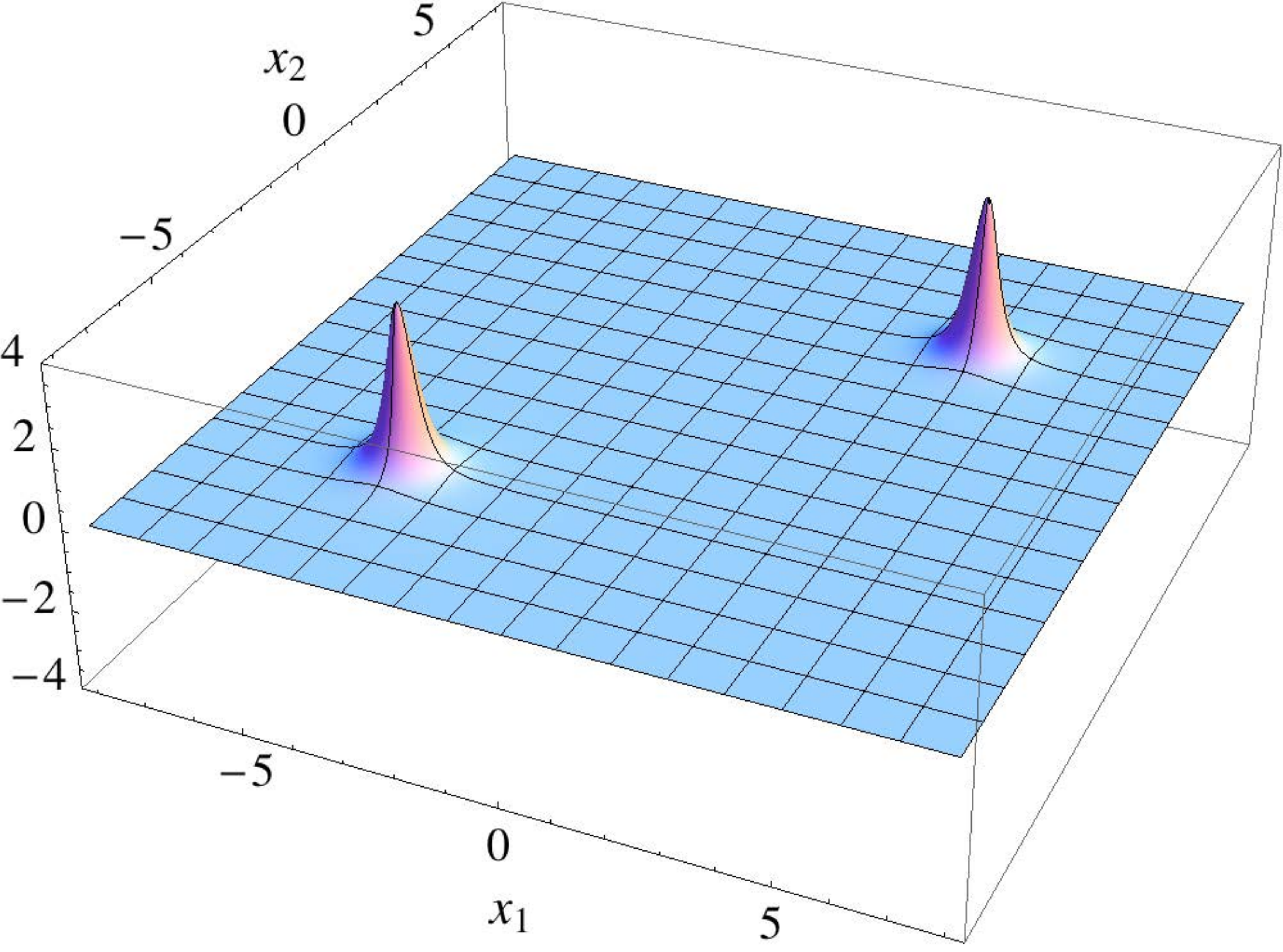} \\ \\
 Action Density of $\omega_{(0)}$: ($S_{(0)}=2$)  & Charge Density of $\omega_{(0)}$: ($Q_{(0)} = 2)$ \\ \\
\includegraphics[width=8cm]{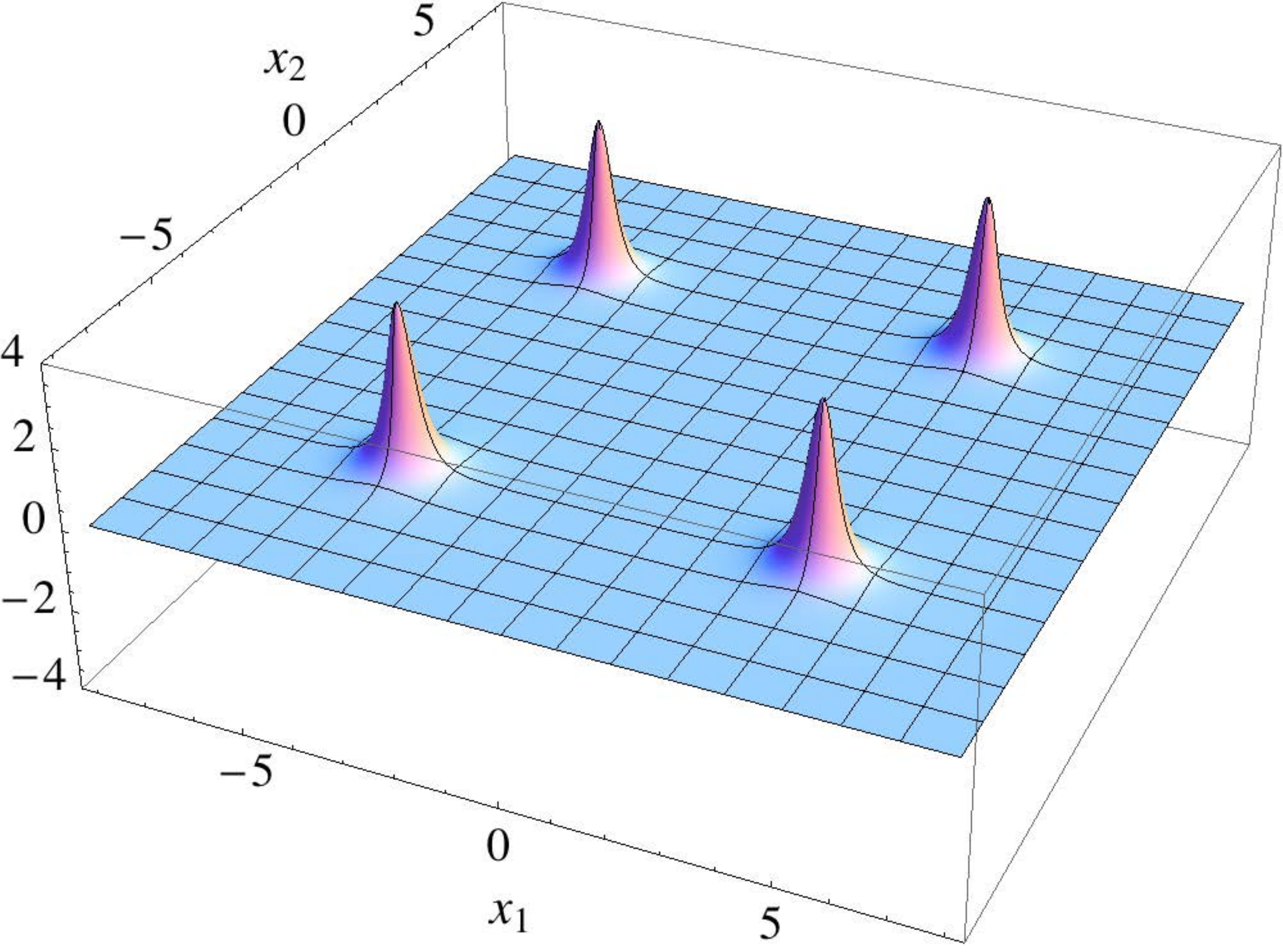} 
 & 
 \includegraphics[width=8cm]{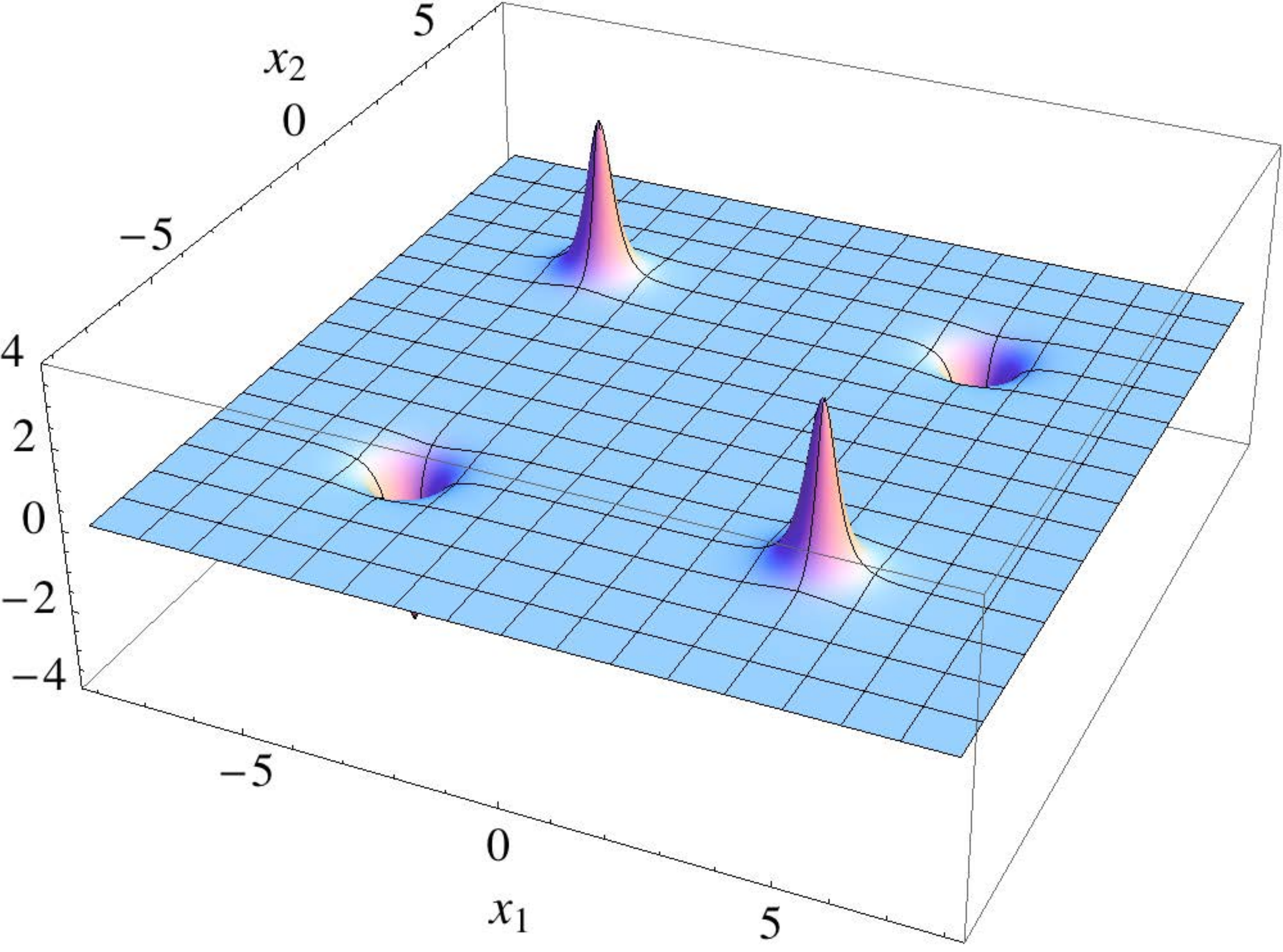} \\ \\
 Action Density of $\omega_{(1)}$: ($S_{(1)}=4$)  & Charge Density of $\omega_{(1)}$: ($Q_{(1)} = 0)$ \\ \\
\includegraphics[width=8cm]{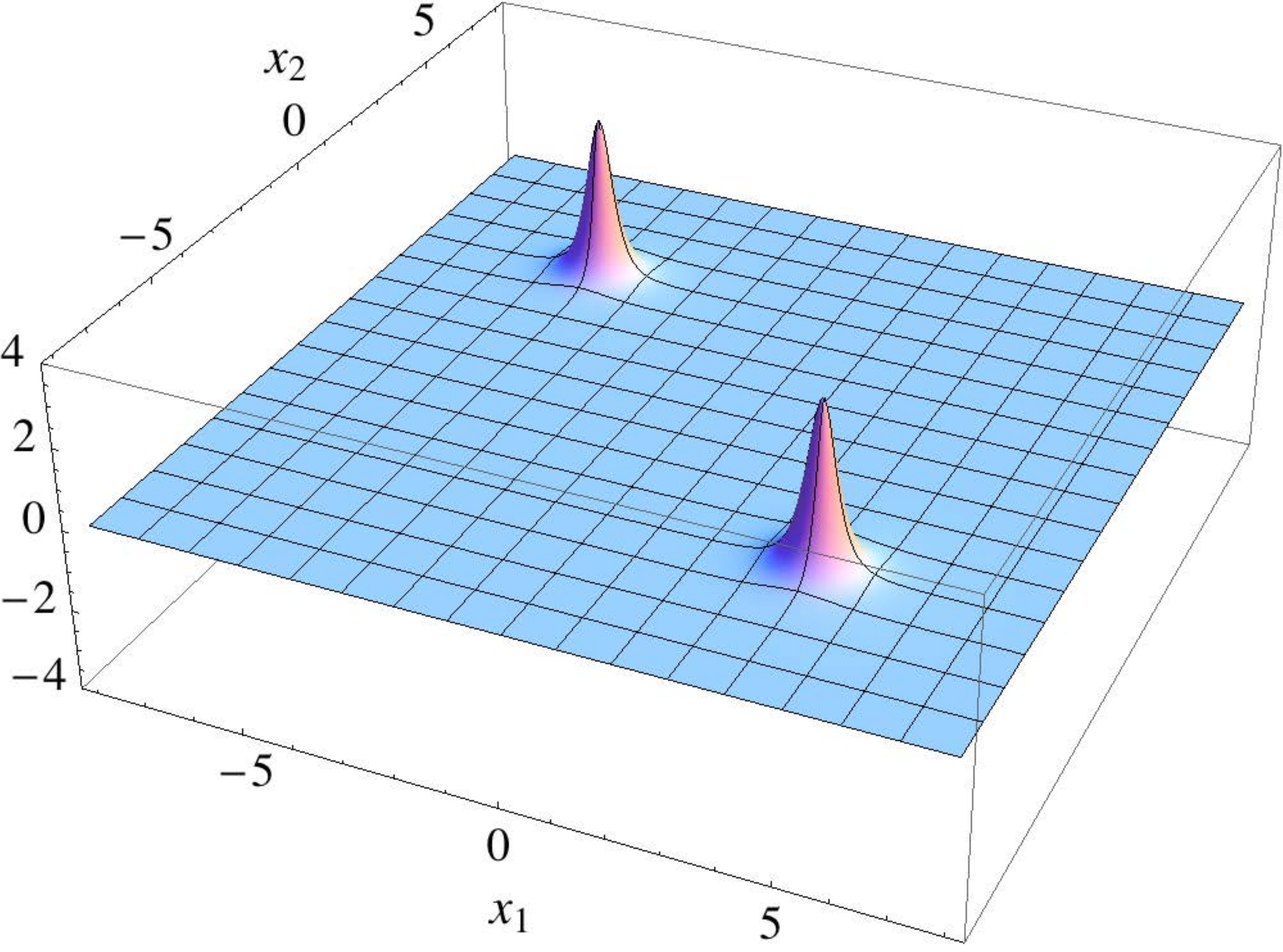} 
 & 
 \includegraphics[width=8cm]{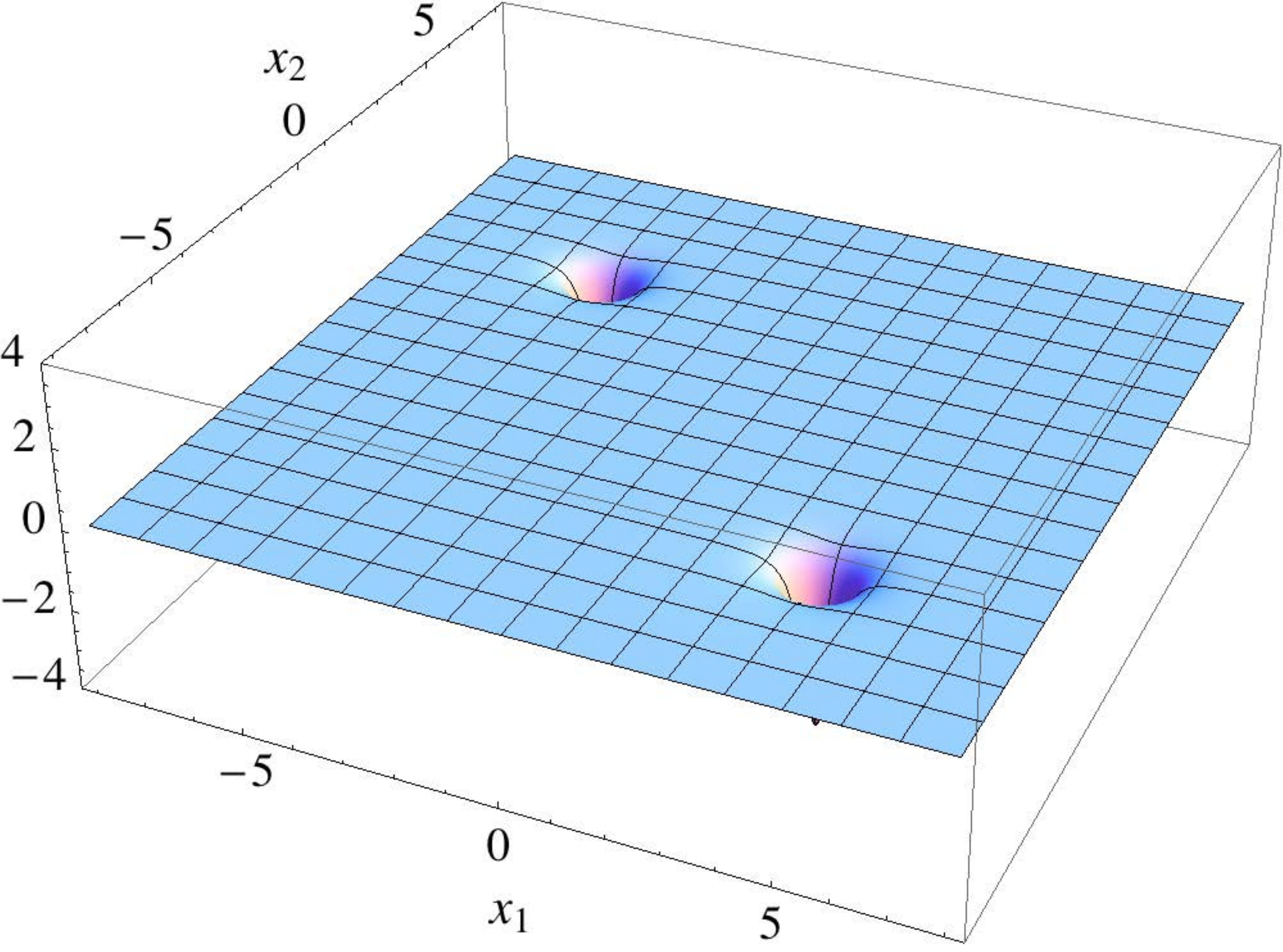} \\ \\
 Action Density of $\omega_{(1)}$: ($S_{(2)}=2$)  & Charge Density of $\omega_{(2)}$: ($Q_{(2)} = -2)$ \\
\end{tabular}
\caption{\label{fig:CP2_R2}The Action and Charge Density configurations due to successive mappings from the ansatz solution (\ref{CP2_R2_ansatz}) in $\mathbb{CP}^2$ on $\mathbb{R}^2$: $\omega_{(0)}=\left(1, \, \lambda \, e^{i \theta_1} \left( z - a \right), \, \mu \, e^{i \theta_2} \left( z^2 - b^2 \right) \right)$ where $a=a_1 + i \, a_2$ and $b=b_1+ i \,b_2$, and plotted for: $\lambda,\mu = 2$, $a_1, a_2 = 0$, $b_1, b_2 = 4$, $\forall$ $\theta_1,\theta_2 \in [0,2\pi)$. The initial configuration $\omega_{(0)}$ corresponds to two instantons, while $\omega_{(1)}$ corresponds to two instantons and two anti-instantons, and $\omega_{(2)}$ corresponds to two anti-instantons. These are all exact solutions to the classical equations of motion, but $\omega_{(1)}$ is non-self-dual.}
\end{figure}

The non-self-dual classical solutions are `unstable'  in the sense that the fluctuation operator about the solution has at least one negative mode. A systematic characterization of the negative modes, and even their number, has not yet been fully performed (see comments in \cite{Din:1980jg,zbook,Jack:1981rq}), but the following physical argument illustrates the point. Consider for example a non-self-dual solution with zero net topological charge, $Q=0$, consisting locally of 2 instantons and 2 anti-instantons. This is the simplest such non-self-dual configuration.  In ${\mathbb C}{\mathbb P}^{N-1}$ a single instanton is characterized by $2N$ parameters, and so has $2N$ zero modes. Therefore, this non-self-dual configuration would have a total of $8N$ zero modes in the infinite separation limit. However, the exact solution at finite separation is constructed by applying projection operators to an initial $Q=2$ instanton, which has just $4N$ zero modes. Thus, the exact non-self-dual solution only has $4N$ zero modes. So, half the zero modes at infinite separation become non-zero-modes, either positive or negative, at finite separation. 
Depending on the parameters, such as orientations, the lifted zero modes may become negative modes or positive modes. As an example, consider the fluctuation
\begin{eqnarray}
n\to \tilde n=n\sqrt{1-\phi^\dagger \phi}+\phi \qquad, \qquad \phi=  D_z n \quad ;\quad  \phi^\dag \cdot n = 0
\label{change}
\end{eqnarray}
for which the change in the action is manifestly negative  \cite{Din:1980jg}:
\begin{eqnarray}
\delta S= -  \int d^2 x \left( \text{Tr}\left[\left(D_z n\right)^\dagger D_{z} n \left( D_{\bar{z}} n\right)^\dag D_{\bar{z}} n \right] +
\text{Tr}\left[ \left(D_{\bar{z}} n \right)^\dagger D_z n \left(D_z n\right)^\dagger D_{\bar{z}} n\right] \right)
\label{unstable}
\end{eqnarray}
In Figure (\ref{fig:potential}) we plot the change in the action as a function of separation, showing how a zero mode at large separation becomes a negative mode at finite separation. This example is for ${\mathbb C}{\mathbb P}^{2}$ ($N=3$) . The action and charge of the corresponding configuration is shown in the second row of Figure (\ref{fig:CP2_R2}).

\section{Explicit Examples on ${\mathbb R}^2$ and ${\mathbb S}^2$}

As Zakrzewski and Din have shown, non-self-dual solutions exist on $\mathbb{R}^2$ (and correspondingly  on compactified $\mathbb{S}^2$)  for the $\mathbb{CP}^{N-1}$ model when $N \ge 3$. These solutions are characterized by a number of parameters that dictate the location, orientation and profile of the configurations and their sub-components. The simplest example occurs for $\mathbb{CP}^2$ $(N=3)$ on $\mathbb{R}^2$, beginning with a two-instanton. This is illustrated in Figure (\ref{fig:CP2_R2}), using the two-instanton ansatz 
\begin{align}
\omega_{(0)}=\left(1, \, \lambda \, e^{i \theta_1} \left( z - a \right), \, \mu \, e^{i \theta_2} ( z^2 - b^2 ) \right)
\label{CP2_R2_ansatz}
\end{align}
This self-dual configuration $\omega_{(0)}$ has total action $S_{(0)}=2$, and total topological charge $Q_{(0)}=2$ (as multiples of $2\pi$), and the parameters $\lambda, \mu > 0$, $a,b \in \mathbb{C}$ and $\theta_1, \theta_2 \in [0,2\pi)$ govern the size, location, and phase orientation of each component single-instanton. After one step, the mapping (\ref{zplus}) produces a non-self-dual configuration $\omega_{(1)}$, whose action and topological charge densities are shown in the second row of Figure (\ref{fig:CP2_R2}). A second projection produces a configuration $\omega_{(2)}$ which is anti-self-dual, comprising two anti-instantons, as shown in the third row of Figure (\ref{fig:CP2_R2}). When $a = 0$, the original solution $\omega_{(0)}$ and mapped solutions, $\omega_{(1)}$ and $\omega_{(2)}$, correspond to symmetric configurations whose individual components are equally spaced, as seen in Figure (\ref{fig:CP2_R2}). The non-self-dual configuration $\omega_{(1)}$ in the second line of Figure (\ref{fig:CP2_R2}) consists of two instantons and two anti-instantons, each of action one,  leading to a total action of $S_{(1)}=4$, and zero total topological charge, $Q_{(1)} = 0$. The final mapping generates a configuration that consists of two anti-instantons of charge $-1$. So we can summarize the action and topological charge values of the tower of solutions as:
\begin{align}
\left( S_{(0)}, Q_{(0)} \right) = \left( 2, 2 \right) \; \xrightarrow[]{\;\; Z_+ \;\;} \;  
\left( S_{(1)}, Q_{(1)} \right)=\left( 4, 0 \right) \; \xrightarrow[]{\;\; Z_+ \;\;} \; 
\left( S_{(2)}, Q_{(2)} \right)=\left( 2, -2 \right)
\end{align}
Note the consistency with the relations in (\ref{n3}).

\begin{figure}[htb]
\begin{tabular}{cc}
\includegraphics[width=8cm]{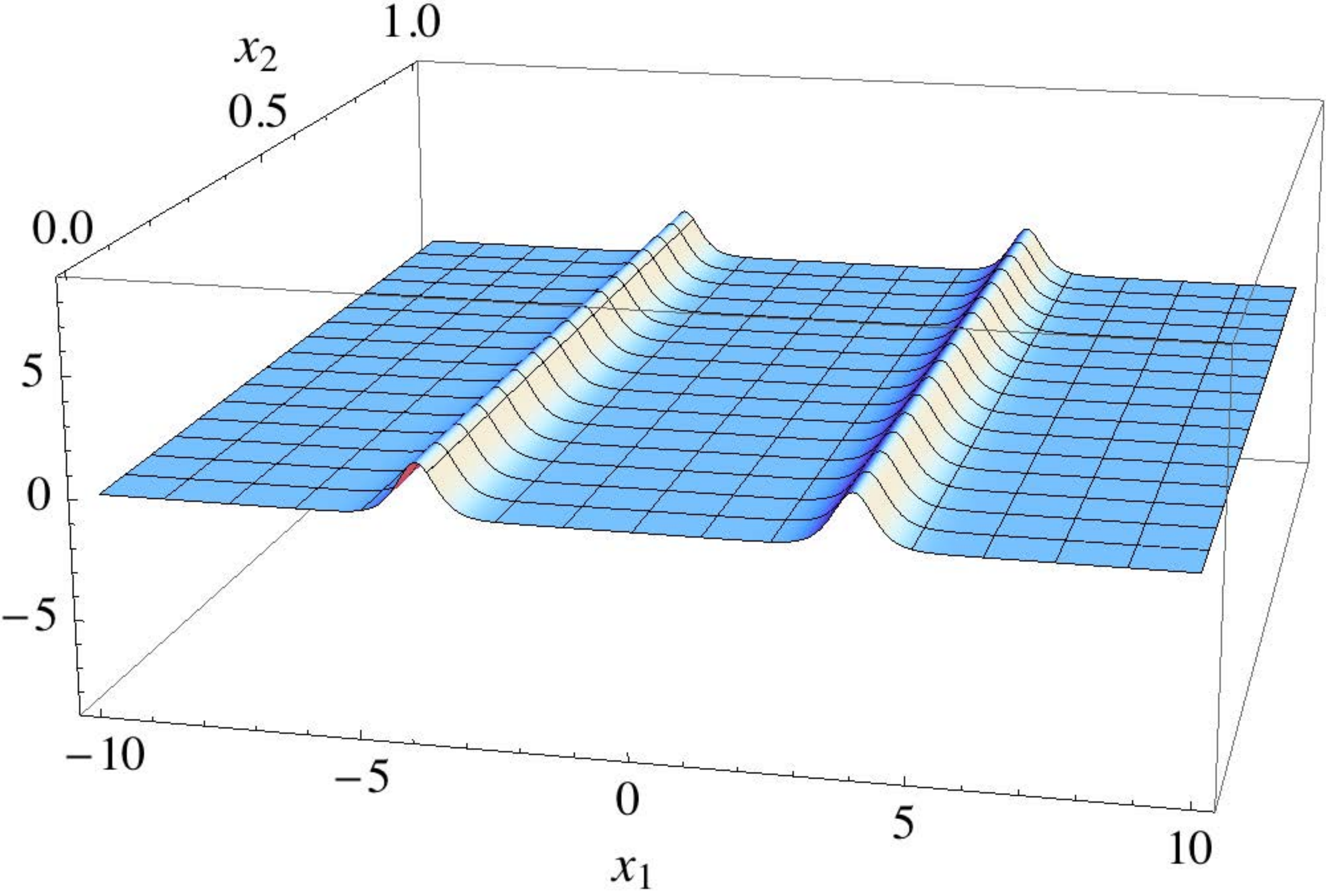} 
 & 
 \includegraphics[width=8cm]{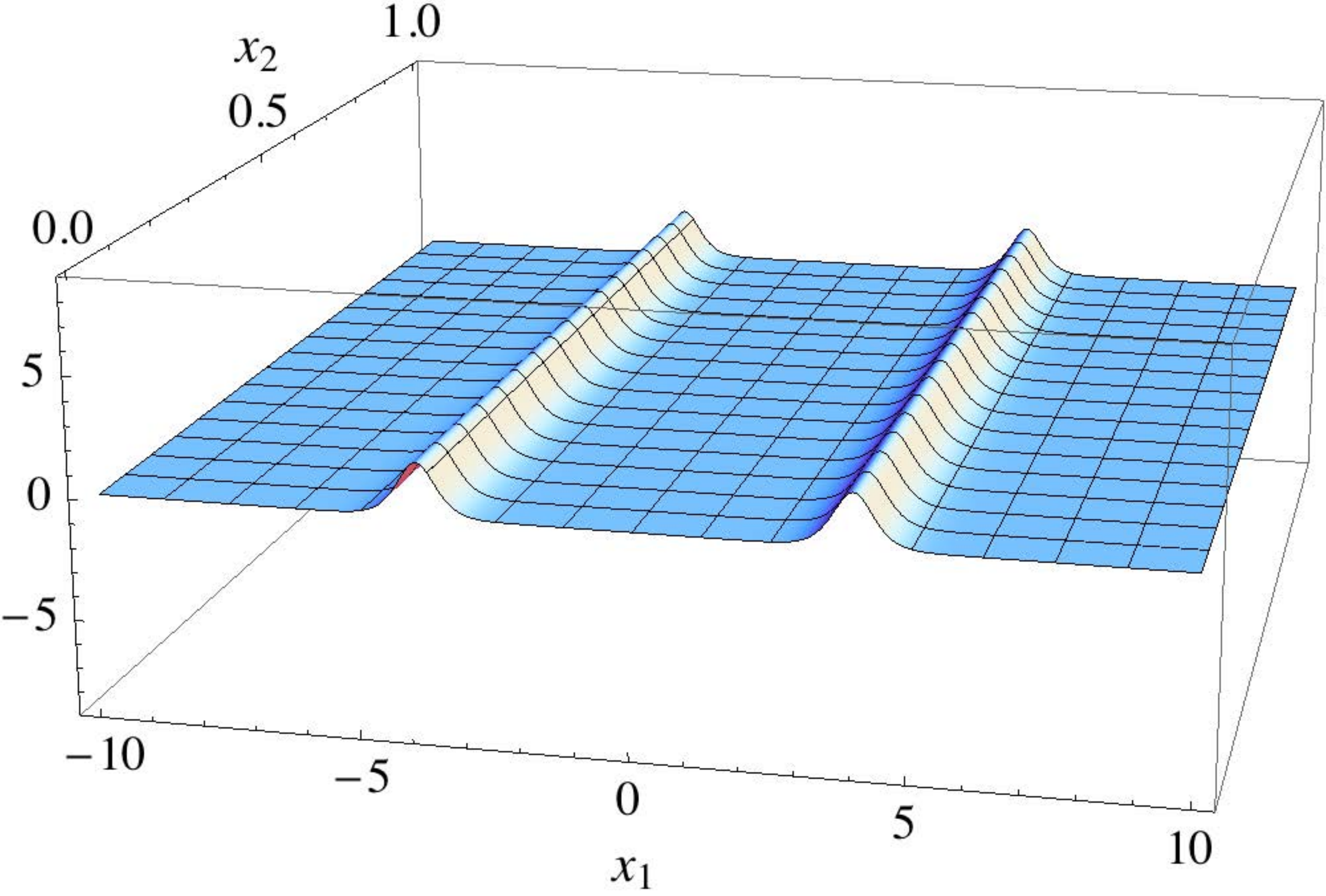} \\ \\
 Action Density of $\omega_{(0)}$: ($S_{(0)}=\frac{2}{3}$)  & Charge Density of $\omega_{(0)}$: ($Q_{(0)} = \frac{2}{3})$ \\ \\
\includegraphics[width=8cm]{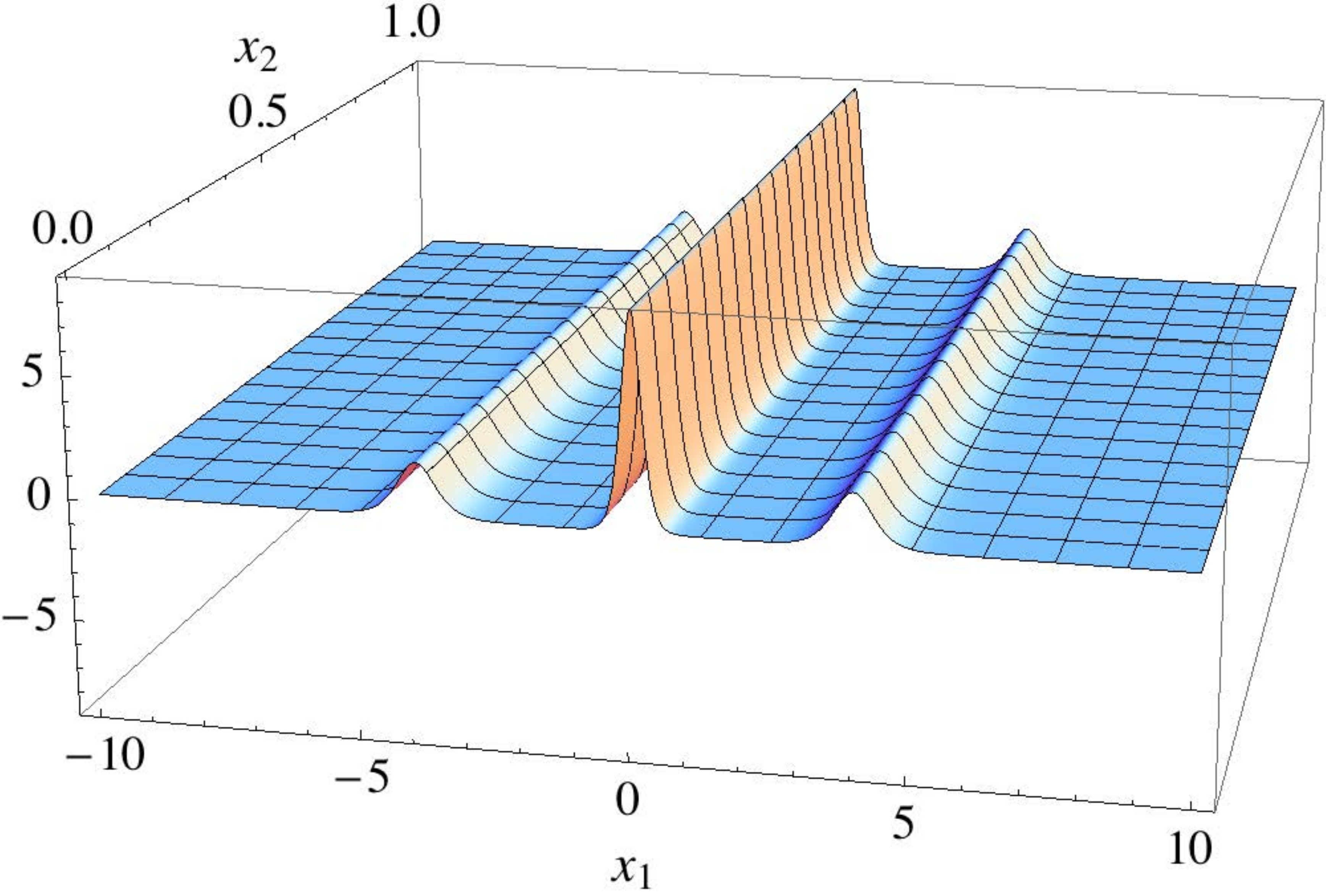} 
 & 
 \includegraphics[width=8cm]{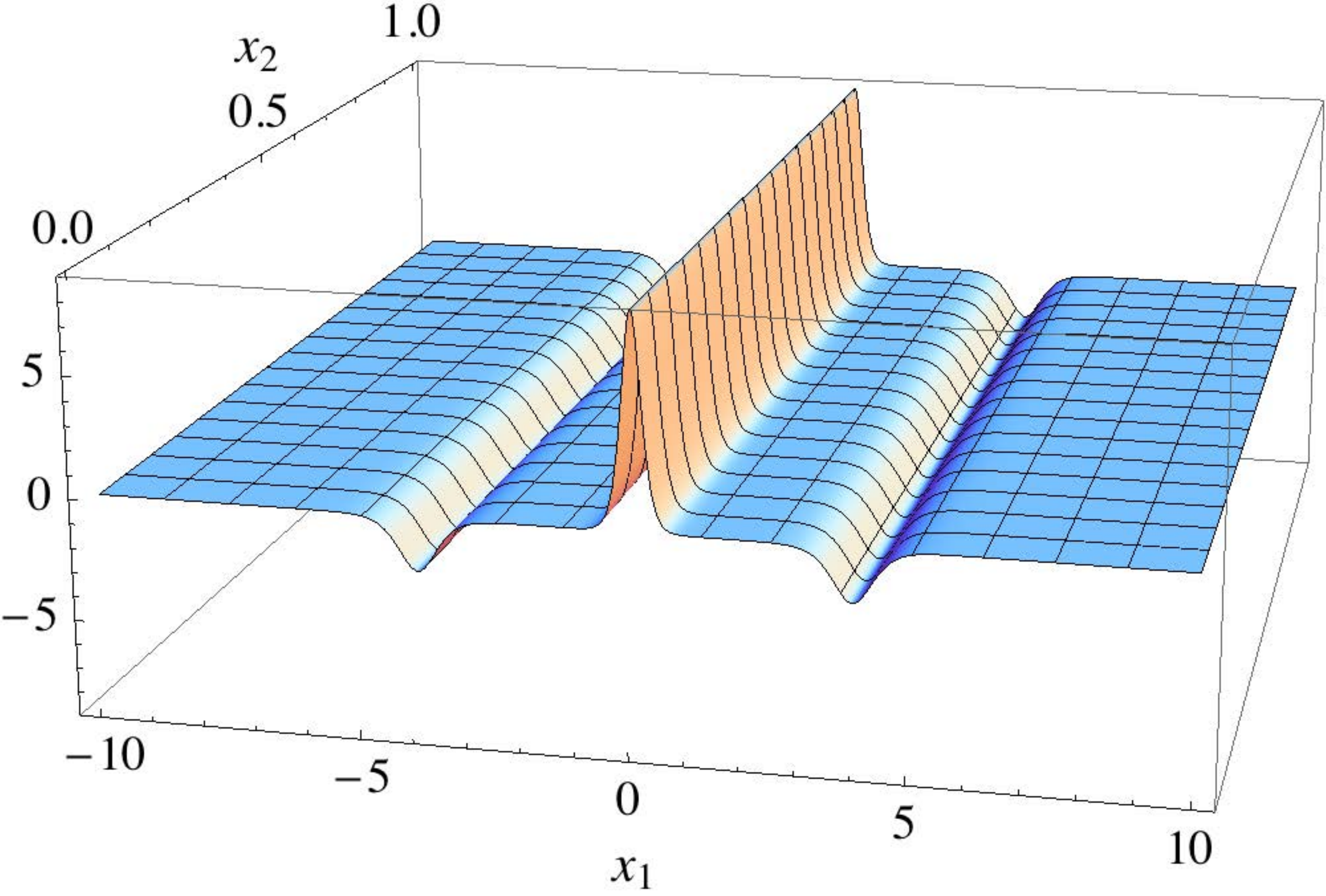} \\ \\
 Action Density of $\omega_{(1)}$: ($S_{(1)}=\frac{4}{3}$)  & Charge Density of $\omega_{(1)}$: ($Q_{(1)} = 0)$ \\ \\
\includegraphics[width=8cm]{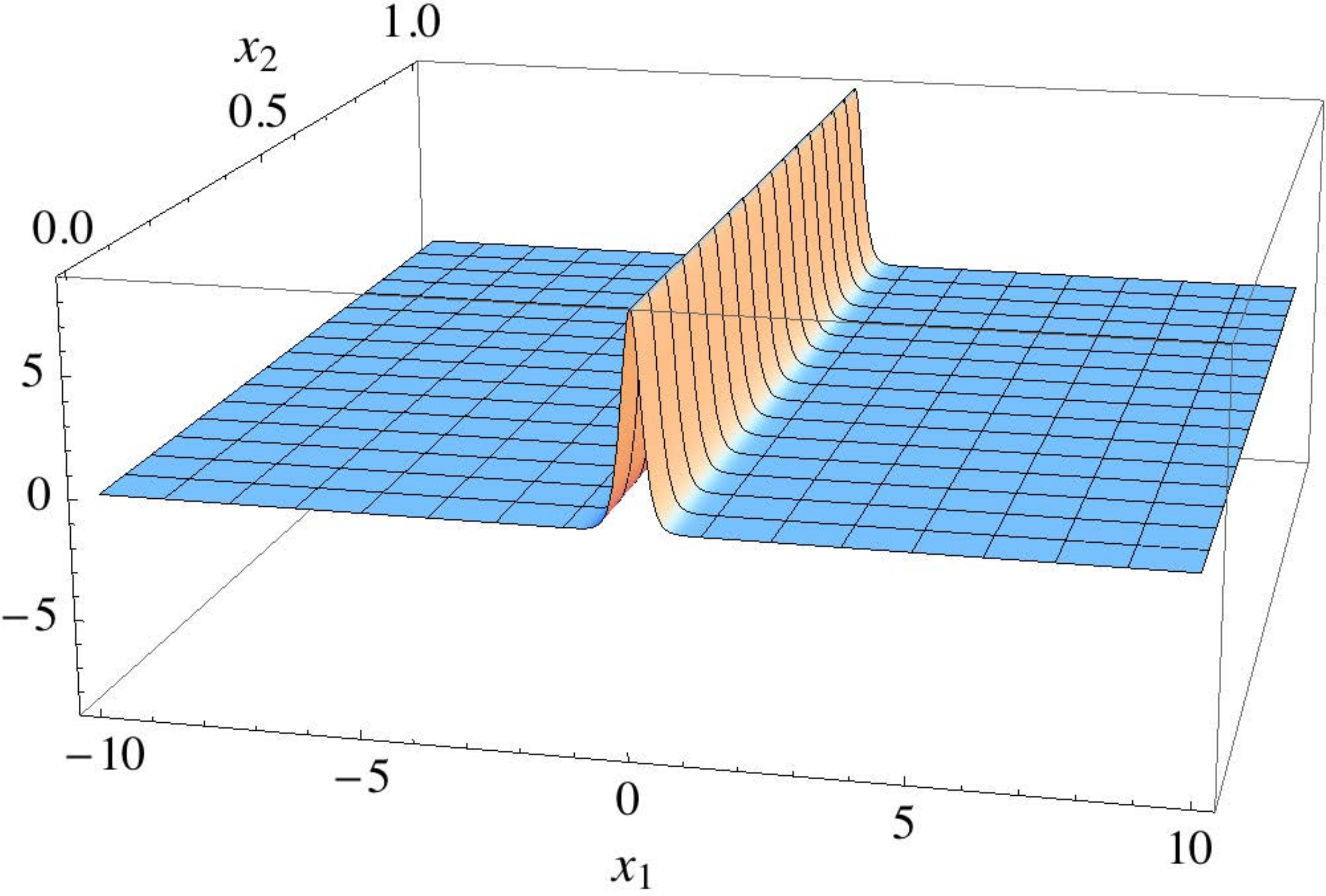} 
 & 
 \includegraphics[width=8cm]{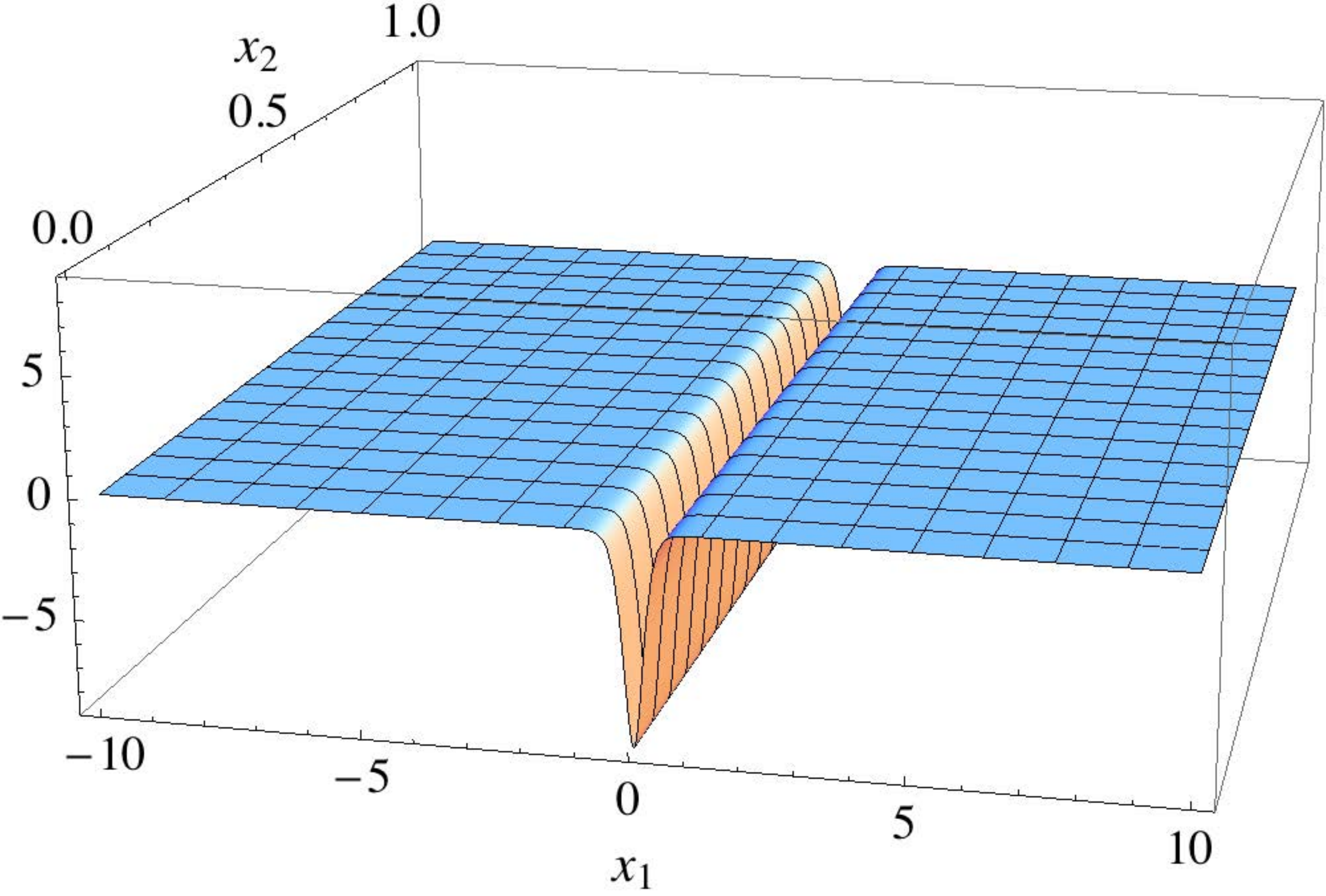} \\ \\
 Action Density of $\omega_{(2)}$: ($S_{(2)}=\frac{2}{3}$)  & Charge Density of $\omega_{(2)}$: ($Q_{(2)} = -\frac{2}{3})$ \\
\end{tabular}
  \caption{The Action and Charge Density configurations due to successive mappings from the ansatz solution (\ref{CP2_Twist_SxR_1_Ansatz}) in $\mathbb{CP}^2$ on $\mathbb{S}_L^1 \times \mathbb{R}^1$:  $\omega_{(0)} = \left( 1, \, \lambda \, e^{i \theta_1} e^{- 2 \pi z /3}, \, \mu \, e^{i \theta_2} e^{- 4\pi z /3}\right)$ where $\lambda = 4000, \mu = 1$, $\forall$ $\theta_1,\theta_2 \in [ 0,2\pi)$. The initial configuration $\omega_{(0)}$ corresponds to two fractionalized instantons each of charge $1/3$, while $\omega_{(1)}$ corresponds to one fractionalized instanton of charge $2/3$ and two fractionalized anti-instantons each of charge $-1/3$, and $\omega_{(2)}$ corresponds to a fractionalized anti-instanton of charge $-2/3$. These are all exact solutions to the classical equations of motion, but $\omega_{(1)}$ is non-self-dual.}
    \label{fig:CP2_Twist_SxR_1}
\end{figure}

\begin{figure}[htb]
\begin{tabular}{cc}
\includegraphics[width=8cm]{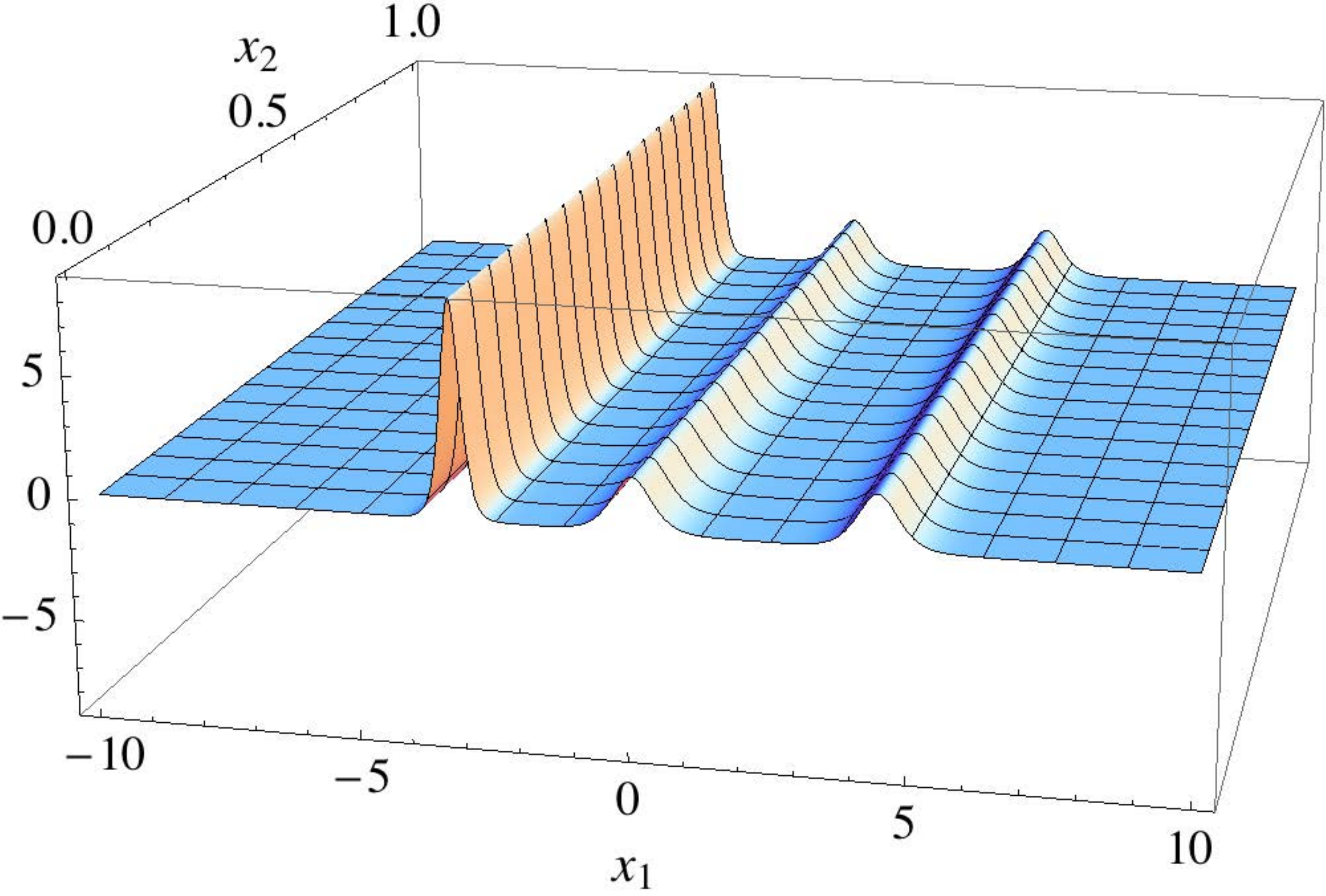} 
 & 
 \includegraphics[width=8cm]{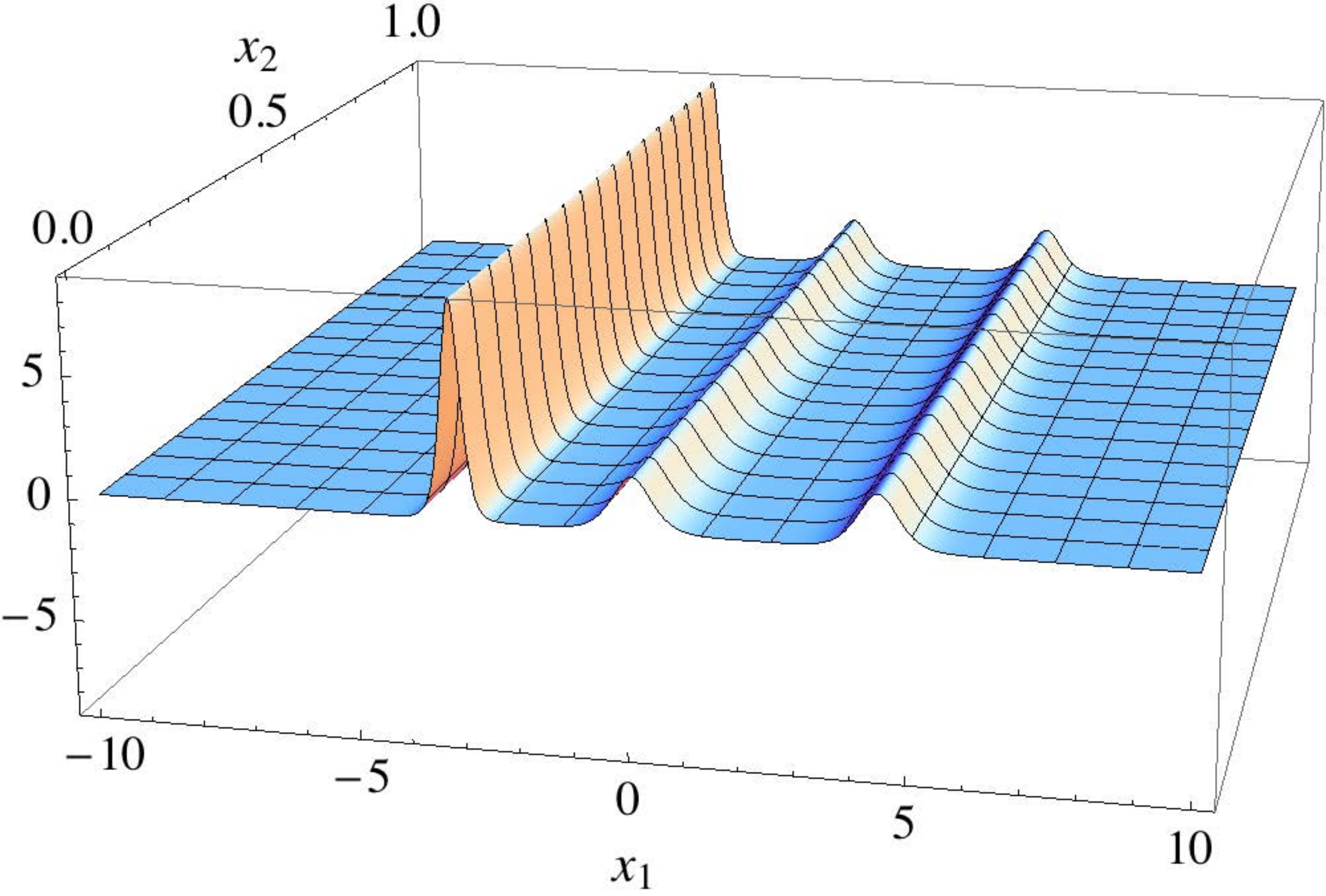} \\ \\
 Action Density of $\omega_{(0)}$: ($S_{(0)}=\frac{4}{3}$)  & Charge Density of $\omega_{(0)}$: ($Q_{(0)} = \frac{4}{3})$ \\ \\
\includegraphics[width=8cm]{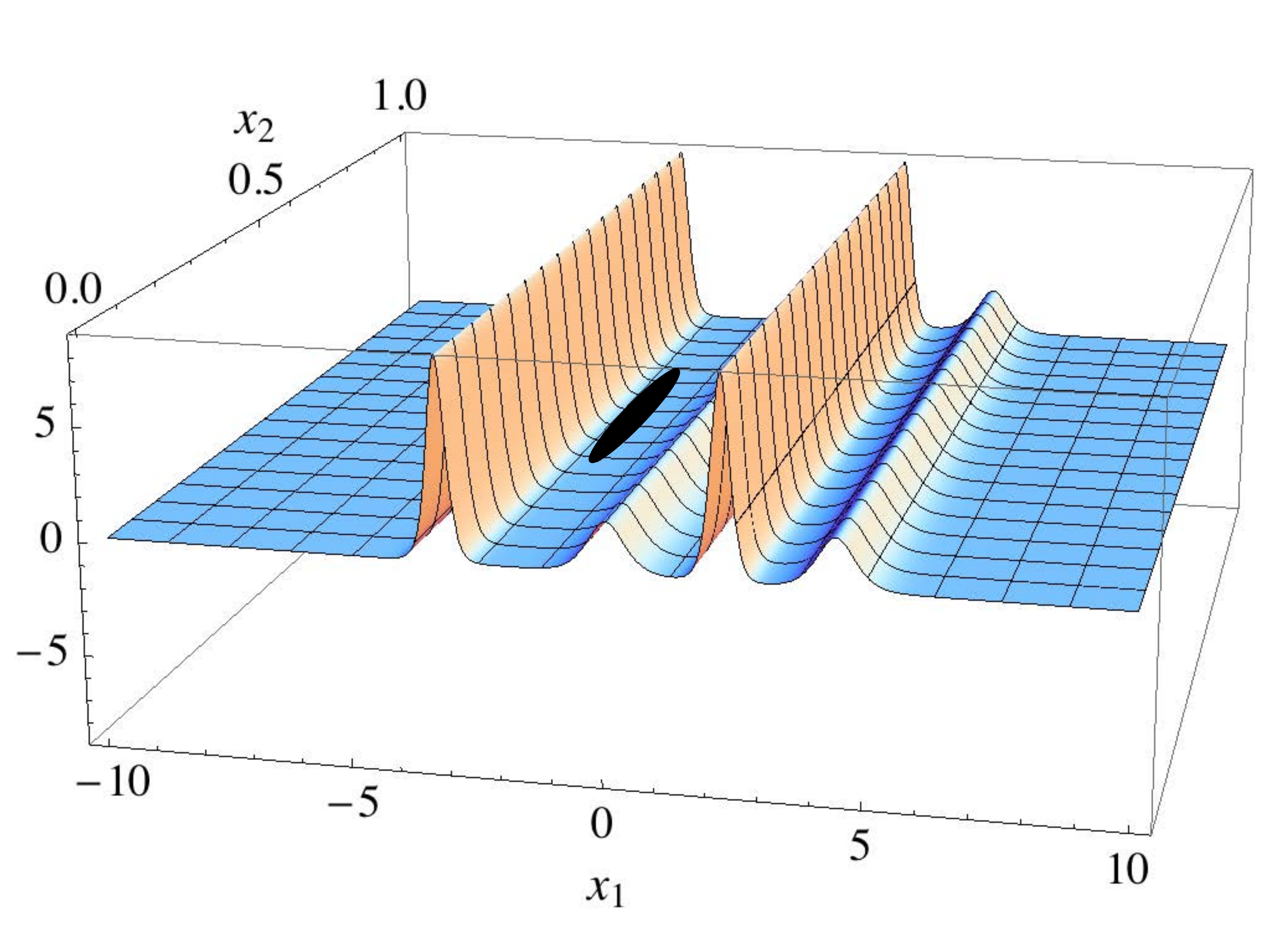} 
 & 
 \includegraphics[width=8cm]{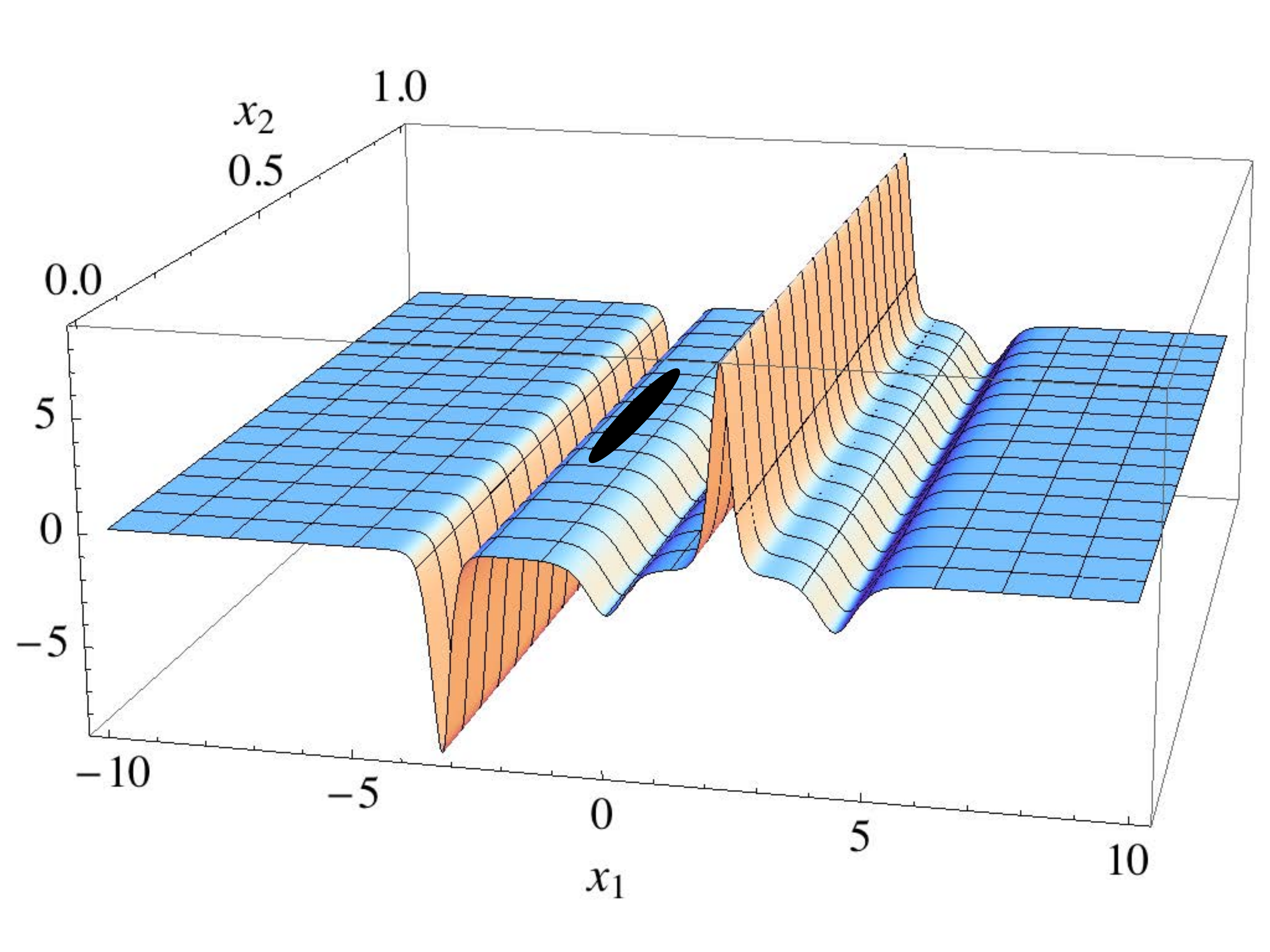} \\ \\
 Action Density of $\omega_{(1)}$: ($S_{(1)}=3$)  & Charge Density of $\omega_{(1)}$: ($Q_{(1)} = \frac{1}{3})$ \\ \\
\includegraphics[width=8cm]{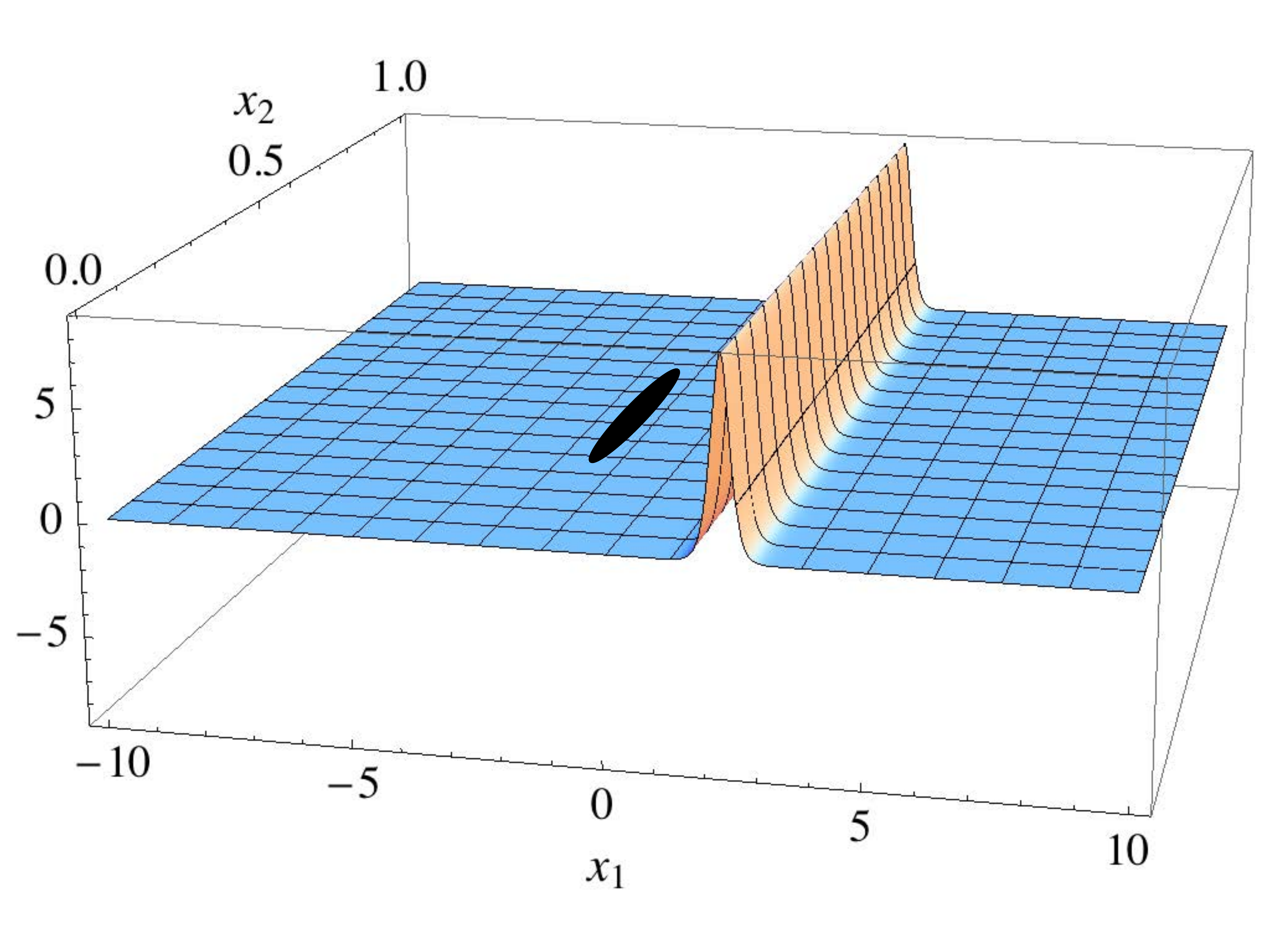} 
 & 
 \includegraphics[width=8cm]{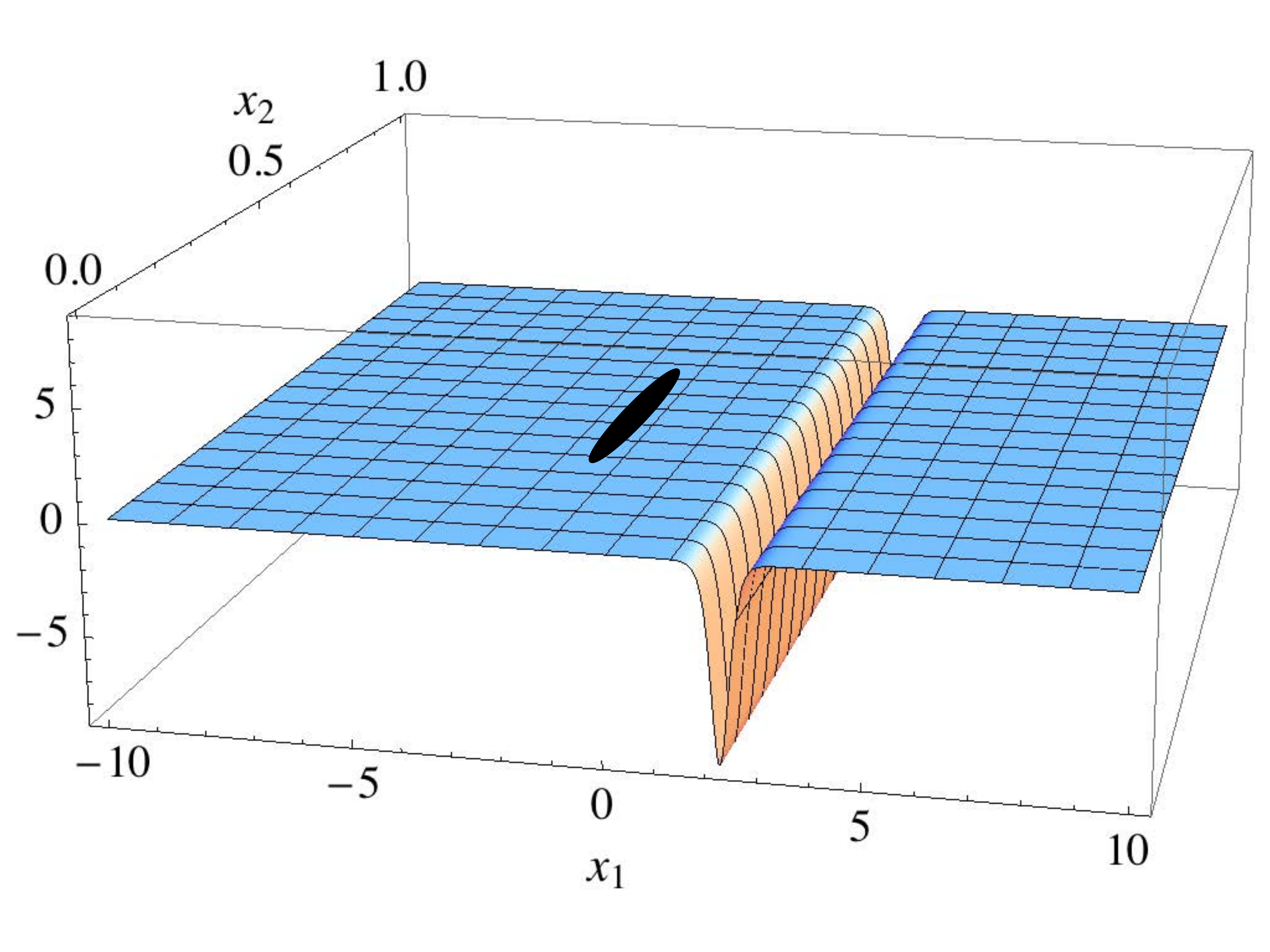} \\ \\
 Action Density of $\omega_{(2)}$: ($S_{(2)}=\frac{5}{3}$)  & Charge Density of $\omega_{(2)}$: ($Q_{(2)} = -\frac{5}{3})$ \\
\end{tabular}
\caption{The Action and Charge Density configurations due to successive mappings from the ansatz solution (\ref{second_ansatz}) in $\mathbb{CP}^2$ on $\mathbb{S}_L^1 \times \mathbb{R}^1$: 
        $\omega_{(0)} = \left( 1, \, \lambda \, e^{i \theta_1} e^{- 2 \pi z /3} + \mu \, e^{i \theta_2} e^{- 8\pi z /3}, \, \nu \, e^{i \theta_3} e^{- 4\pi z /3}\right)$ where $\lambda=10^4, \mu=10^{-2}, \nu = 10^4$, $\theta_1 = \pi, \theta_2 = 0$, $\forall$ $\theta_3 \in [ 0,2\pi)$. The initial configuration $\omega_{(0)}$ corresponds to two fractionalized instantons each of charge $1/3$ and another fractionalized instanton of charge $2/3$, while $\omega_{(1)}$ corresponds to one instanton of charge $2/3$ and another of charge $1$ (marked by the black oval) and two anti-instantons each of charge $-1/3$ and another anti-instanton of charge $-2/3$, and $\omega_{(2)}$ corresponds to an anti-instanton of charge $-2/3$ and an anti-instanton of charge $-1$ (marked by the black oval).  Notice the appearance of very sharp instanton and anti-instanton peaks in the third, fourth, fifth and sixth plots, marked by the black oval shape, as discussed in the text. These peaks are so sharp that they do not show up on the same scale, but their cross-sections are plotted in Figure  \ref{fig:HighlocalInstanton}. Note that $\omega_{(0)}$, $\omega_{(1)}$ and $\omega_{(2)}$ are all exact solutions to the classical equations of motion, but $\omega_{(1)}$ is non-self-dual.}
        \label{fig:CP2_Twist_SxR_60}
\end{figure}

\begin{figure}[htb]
\begin{tabular}{cc}
\includegraphics[width=8cm]{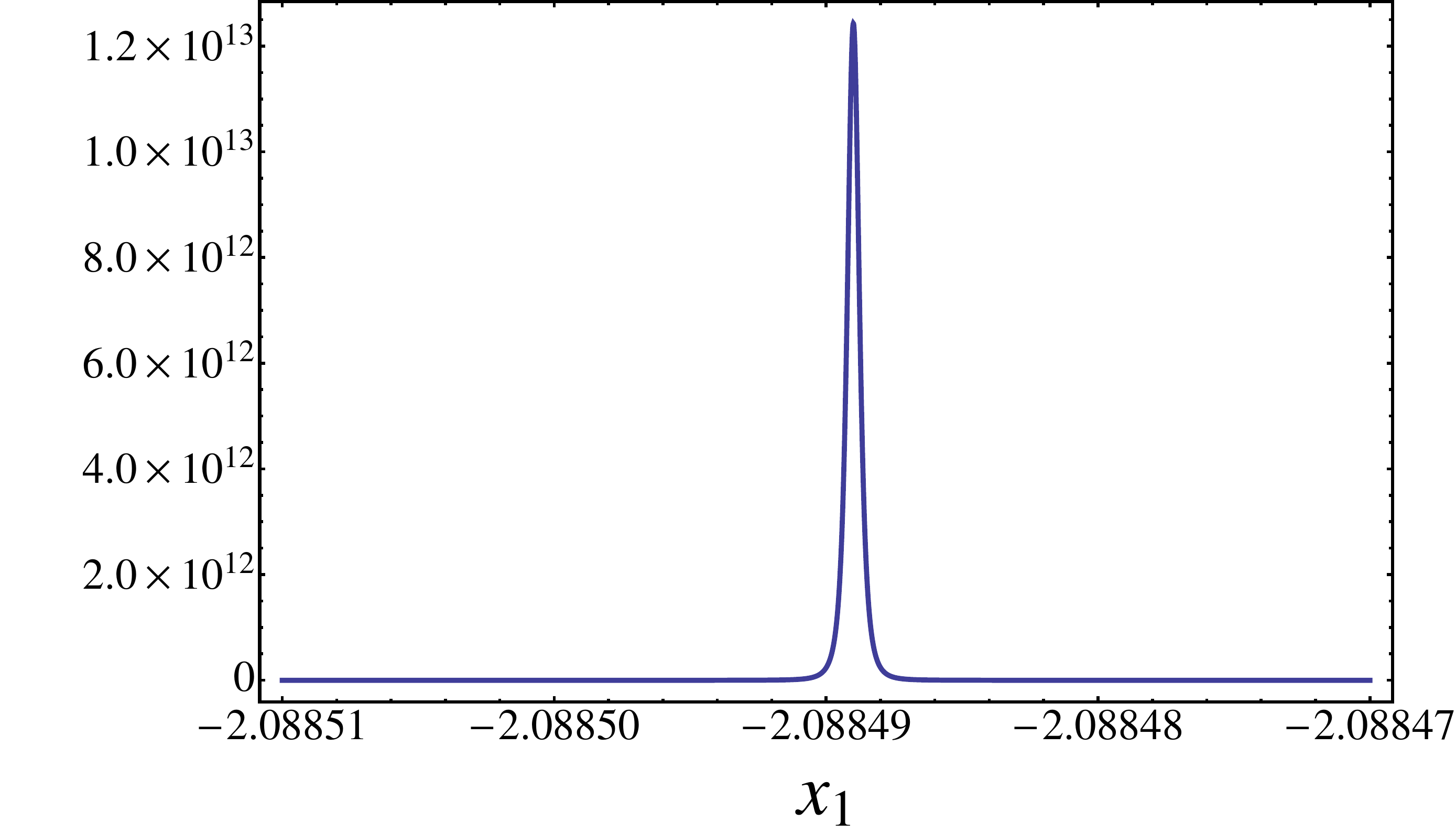}
&
\includegraphics[width=8cm]{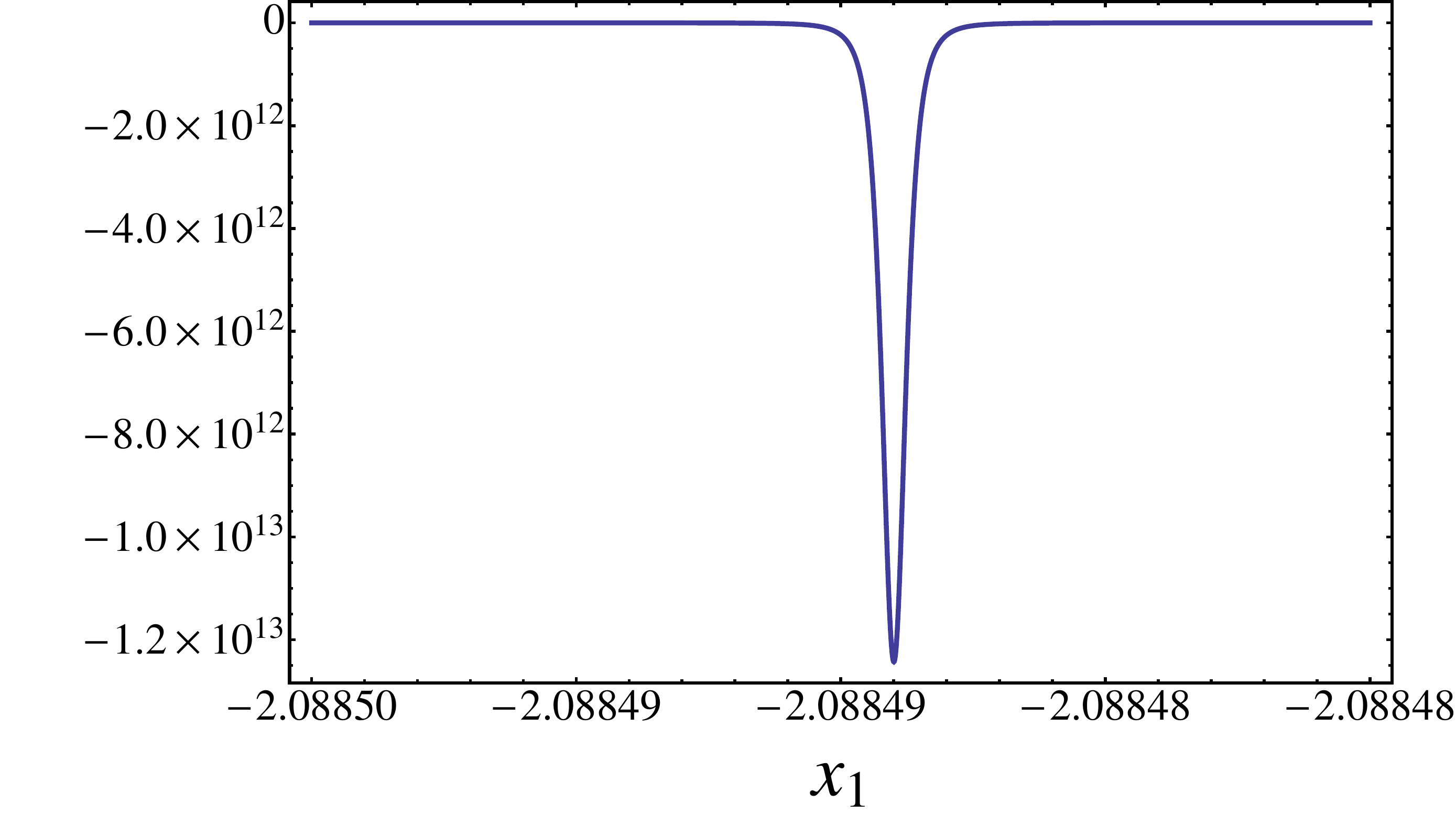}
\end{tabular}
\caption{A magnified cross section of the charge density of the highly localized charge 1 instanton and anti-instanton that appear in the fourth and sixth plots in Figure (\ref{fig:CP2_Twist_SxR_60}). Both are plotted with the same parameters used in Figure (\ref{fig:CP2_Twist_SxR_60}).}
\label{fig:HighlocalInstanton}
\end{figure}

\section{Explicit Examples on  ${\mathbb S}^1_L \times{\mathbb R}^1$}

As in \cite{Dunne:2012ae,Dunne:2012zk}, we impose ${\mathbb Z}_N$ twisted boundary conditions in the compactified spatial direction:
\begin{eqnarray}
n(x_1, x_2+L)=\Omega \, n(x_1, x_2)\qquad, \qquad \Omega = {\rm diag}\left(1, e^{-2\pi i/N}, \dots, e^{-2\pi i (N-1)/N}\right)
\label{twisted}
\end{eqnarray}
This corresponds to the same condition on the homogeneous field $\omega(x_1, x_2)$, and  we see from (\ref{zplus}) that if the initial instanton solution $\omega_{(0)}$ satisfies ${\mathbb Z}_N$ twisted boundary conditions, then all subsequent projected solutions in (\ref{zplus}), (in particular, the non-self-dual ones), also satisfy ${\mathbb Z}_N$ twisted boundary conditions. 

For self-dual solutions, the fractionalization arises because of an interplay between the twisted boundary condition, which could be imposed by phase factors in the compactified $x_2$ direction, 
and the holomorphicity condition for an instanton. Thus for an instanton, the twists must arise from factors expressed in terms of the holomorphic variable $z=x_1+i x_2$, and so the twists in the compact $x_2$ direction necessarily also affect the form of the solution in the non-compact $x_1$ direction \cite{Bruckmann:2007zh,Brendel:2009mp,Dunne:2012ae,Dunne:2012zk}. For non-self-dual solutions, the fractionalization is inherited from the fractionalization of the initial self-dual solution $\omega_{(0)}$.

We illustrate the effect of twisted boundary conditions on some non-self-dual configurations in $\mathbb{CP}^2$, for which $N=3$. The first example demonstrates a configuration analogous  to that in Figure (\ref{fig:CP2_R2}) on $\mathbb{R}^2$, while the second demonstrates a new effect not seen on $\mathbb{R}^2$.
These examples also serve to demonstrate the diversity of non-self-dual solutions that are possible with twisted boundary conditions in $\mathbb{CP}^2$ on $\mathbb{S}_L^1 \times \mathbb{R}^1$.

\underline{Example 1}: Figure (\ref{fig:CP2_Twist_SxR_1}) shows the simplest non-self-dual solution, manifest in $\mathbb{CP}^2$, on $\mathbb{S}_L^1 \times \mathbb{R}^1$ with $\mathbb{Z}_3$ twisted boundary conditions. We take an initial two-instanton ansatz
\begin{align}
\omega_{(0)} = \left( 1, \, \lambda \, e^{i \theta_1} e^{-2 \pi z / 3}, \, \mu \, e^{i \theta_2} e^{-4 \pi z / 3} \right) ,
\label{CP2_Twist_SxR_1_Ansatz}
\end{align}
where $\lambda,\mu > 0$, $\theta_1,\theta_2 \in [ 0 , 2 \pi)$. This solution is self-dual, with action and charge $S_{(0)}=Q_{(0)} = 2/3$, consisting of two separate fractionalized instantons of charge $1/3$. After one application of the mapping (\ref{zplus}) we obtain a non-self-dual configuration $\omega_{(1)}$ with zero net topological charge $Q_{(1)} = 0$, and action $S_{(1)}=4/3$, as shown in the second row of Figure (\ref{fig:CP2_Twist_SxR_1}). We can identify this configuration as consisting of a double-instanton of charge $2/3$ at the midpoint of the original instanton components, with two anti-instantons each of charge $-1/3$, located near the positions of the original instanton components. Note the difference from the example on $\mathbb{R}^2$ in Figure~(\ref{fig:CP2_R2}).
After one  further application of the mapping (\ref{zplus}) we obtain an anti-self-dual configuration $\omega_{(2)}$, which for this choice of parameters looks like a double (fractionalized)  anti-instanton configuration, with total charge $-2/3$.
So we can summarize the action and topological charge values of the tower of  solutions as:
\begin{align}
\left( S_{(0)}, Q_{(0)} \right) =  \left( \frac{2}{3}\,, \frac{2}{3} \right) \; \xrightarrow[]{\;\; Z_+ \;\;} \;  
\left( S_{(1)}, Q_{(1)} \right)=\left( \frac{4}{3}\,, 0 \right)  \; \xrightarrow[]{\;\; Z_+ \;\;} \; 
\left( S_{(2)}, Q_{(2)} \right)=\left( \frac{2}{3}\,, - \frac{2}{3} \right)
\end{align}
Note the consistency with the relations in (\ref{n3}).
\\

\underline{Example 2}:  Figure (\ref{fig:CP2_Twist_SxR_60}) shows a non-trivial non-self-dual solution in $\mathbb{CP}^2$ on $\mathbb{S}_L^1 \times \mathbb{R}^1$ with $\mathbb{Z}_3$ twisted boundary conditions. We begin with the initial instanton ansatz
\begin{align}
\omega_{(0)} = \left( 1, \, \lambda \, e^{i \theta_1} e^{- 2 \pi z /3} + \mu \, e^{i \theta_2} e^{- 8\pi z /3}, \, \nu \, e^{i \theta_3} e^{- 4\pi z /3}\right) ,
\label{second_ansatz}
\end{align}
where $\lambda,\mu,\nu > 0$, $\theta_1,\theta_2,\theta_3 \in [0,2\pi)$. This starting configuration consists of three instantons of topological charge $2/3,1/3$ and $1/3$, respectively, producing $S_{(0)}=Q_{(0)} = 4/3$. On comparison with (\ref{CP2_Twist_SxR_1_Ansatz}) and Figure (\ref{fig:CP2_Twist_SxR_1}), we note that the inclusion of the extra $\mathbb{Z}_3$ twist preserving term $\text{exp}\!\left[ - 8 \pi z / 3 \right]$ directly contributes the extra charge $2/3$ instanton in the starting configuration, and imbues greater structure to the subsequent non-self-dual solution. 
At first sight, the non-self-dual configuration $\omega_{(1)}$ plotted in the second row of Figure (\ref{fig:CP2_Twist_SxR_60}) appears to consist of one instanton of charge $2/3$, and three anti-instantons, two of charge $-1/3$, and one of charge $-2/3$. This would suggest a net charge of $-2/3$ and net action equal to $2$. However, there is another instanton, of net charge $+1$, which for these parameters is not fractionalized, that is a very sharp peak that can not be seen on the scale of the figure. It is marked by the black oval in the plots in the second line of Figure (\ref{fig:CP2_Twist_SxR_60}). A magnified cross-section of this extra instanton is shown in Figure \ref{fig:HighlocalInstanton}. Thus, the actual assignment of action and charge, which is easily verified by numerical integration, is $S_{(1)}=3$, and $Q_{(1)} = 1/3$.
It is interesting to see that for this non-self-dual configuration some sub-components are clearly fractionalized, while there is a distinct lump that is not. 
This demonstrates a richer structure when compared to the non-self-dual solutions on $\mathbb{R}^2$ (and $\mathbb{S}^2$). 
The unresolved sharp peaks in the configurations of Figure (\ref{fig:CP2_Twist_SxR_60}), marked by a black oval shape, correspond to this highly localized non-fractionalized instanton and anti-instanton, and have the resolved form shown in Figure (\ref{fig:HighlocalInstanton}).
In addition, the extra twist preserving term also affects the final mapped configuration $\omega_{(2)}$, with further structure when compared to Figure (\ref{fig:CP2_Twist_SxR_1}). Thus, while at first sight, it looks like the final configuration $\omega_{(2)}$ has $S_{(2)}=-Q_{(2)} = 2/3$, in fact there is a very sharply peaked anti-instanton at the location marked by the black oval, leading to  the net result: $S_{(2)}=-Q_{(2)} = 5/3$. Observe that, unlike the $\mathbb{R}^2$ examples,  the total action and topological charge of the final solution $\omega_{(0)}$ need not be equal to those of the starting solution $\omega_{(2)}$. 
So we summarize the action and topological charge values of this tower of  solutions as:
\begin{align}
\left( S_{(0)}, Q_{(0)} \right) =  \left( \frac{4}{3}\, , \frac{4}{3} \right) \; \xrightarrow[]{\;\; Z_+ \;\;} \;  
\left( S_{(1)}, Q_{(1)} \right)=\left( 3\, , \frac{1}{3} \right)  \; \xrightarrow[]{\;\; Z_+ \;\;} \; 
\left( S_{(2)}, Q_{(2)} \right)=\left( \frac{5}{3}\, , - \frac{5}{3} \right)
\end{align}
Note again  the consistency with the relations in (\ref{n3}).
With the inclusion of further twist preserving terms, an even richer set of solutions develop that are unique to twisted boundary conditions on $\mathbb{S}_L^1 \times \mathbb{R}^1$, generating all multiples of $1/3$ for the charge of the non-self-dual solution $\omega_{(1)}$.

\section{Conclusion}

In this paper we have shown that Din and Zakrzewski's construction of non-self-dual classical solutions in the ${\mathbb C}{\mathbb P}^{N-1}$ model on ${\mathbb R}^2$ and ${\mathbb S}^2$
extends naturally to non-self-dual classical solutions on ${\mathbb S}^1_L \times{\mathbb R}^1$, with ${\mathbb Z}_N$ twisted boundary conditions. As occurs for the self-dual instantons, the non-self-dual solutions  fractionalize into sub-component objects, which we can identity locally as fractionalized instantons and anti-instantons. This leads to a much richer spectrum of actions and charges, generically in integer units of $1/N$ for ${\mathbb C}{\mathbb P}^{N-1}$. We furthermore propose that the physical significance of these `unstable' non-self-dual solutions is not associated with unstable vacuum decay, but rather that in a semi-classical saddle point analysis of the path integral they produce imaginary non-perturbative terms that match (and cancel against) imaginary non-perturbative terms arising in the perturbative sector due to the infrared-renormalon-induced non-Borel-summability of perturbation theory for ${\mathbb C}{\mathbb P}^{N-1}$. This suggests that it would be worthwhile to classify and analyze more systematically the negative modes corresponding to these exact non-self-dual solutions. Technically, in ${\mathbb C}{\mathbb P}^{N-1}$ we see that these negative modes arise as some would-be zero-modes associated with an approximate non-self-dual configuration of infinitely-far-separated instantons and anti-instantons, become negative modes as these sub-components approach one another;  the exact non-self-dual solution has fewer zero-modes  than its sub-components would suggest, because it inherits these zero-modes from the parameters of the simpler initial self-dual configuration $\omega_{(0)}$. We expect similar behavior in twisted Yang-Mills theory, although the 
${\mathbb C}{\mathbb P}^{N-1}$ case is simpler and more explicit. Finally, we note that similar effects should also occur in other 2d sigma models.

\section{Appendix}
In this Appendix we list some useful identities concerning the non-self-dual configurations generated by the mapping (\ref{tower}).
For all classical solutions generated by (\ref{tower}), we have:
\begin{subequations}
\begin{align}
\omega_{(k)}^\dag \, \omega_{(l)} &= 0 \quad \text{if} \;\; k \ne l																\\
\partial_{\bar{z}} \, \omega_{(k)} &= {}- \omega_{(k-1)} \, \frac{| \omega_{(k)} |^2}{| \omega_{(k-1)} |^2}									\\
\partial_z \! \left( \frac{\omega_{(k-1)}}{| \omega_{(k-1)} |^2} \right) &= \frac{\omega_{(k)}}{| \omega_{(k-1)} |^2}							\\
\omega_{(N)} &= Z_+ \omega_{(N-1)} = 0 \label{MappingOverCondition00}
\end{align}
\end{subequations}

In terms of the projectors:
\begin{align}
\omega_{(k+1)} \propto \partial_z \mathbb{P}_{(k)} \omega_{(k)} \qquad, \qquad \omega_{(k-1)} \propto \partial_{\bar{z}} \mathbb{P}_{(k)} \omega_{(k)}
\label{altprojections}
\end{align}
The following projector identities are useful in determining (\ref{sumid}), and are general for all Grassmanians:
\begin{subequations}
\begin{align}
\mathbb{P}_{(i)} \, \mathbb{P}_{(j)} &= \mathbb{P}_{(i)} \delta_{ij} \\
\mathbb{P}_{(i)} \, \partial_z \mathbb{P}_{(i)} \, \mathbb{P}_{(i)} &= 0 \quad \forall \, i \\
\mathbb{P}_{(i)} \, \partial_z \mathbb{P}_{(j)} &= 0 \quad \text{if} \quad j = i +1 \quad \text{or} \quad | i - j | \ge 2 \\
\partial_z \mathbb{P}_{(i)} \, \mathbb{P}_{(j)} &= 0 \quad \text{if} \quad j = i +1 \quad \text{or} \quad | i - j | \ge 2 \\
\partial_z \mathbb{P}_{(i)} \, \partial_z \mathbb{P}_{(j)} &= 0 \quad \text{if} \quad j = i + 1 \quad \text{or} \quad j = i + 2 \quad \text{or} \quad | i - j | \ge 3  \\
\partial_z \mathbb{P}_{(i)} \, \partial_{\bar{z}} \mathbb{P}_{(j)} &= 0 \quad \text{if} \quad | i - j | \ge 2
\end{align}
\end{subequations}
Additional identities are found by taking the Hermitian conjugate since $\left( \partial_z \mathbb{P}_{(i)}\right)^\dag = \partial_{\bar{z}} \mathbb{P}_{(i)}$.

\section*{Acknowledgement} 

We acknowledge support from DOE grant DE-FG02-92ER40716. GD thanks M. \"Unsal and A. Yung for discussions.

\end{document}